\documentclass[accepted]{uai2021} 

\usepackage[american]{babel}

\usepackage{natbib} 
\usepackage{mathtools} 
\usepackage{booktabs} 
\usepackage{tikz} 

\usepackage{amsmath,amssymb,amsfonts}
\usepackage{algorithmic}
\usepackage{graphicx}
\usepackage{textcomp}
\usepackage{bmpsize}
\usepackage{xcolor}
\usepackage{lipsum}

\usepackage{slashbox}
\usepackage{multirow}
\usepackage{comment}
\usepackage{amsthm}
\usepackage{bm}

\usepackage{amsmath,amssymb}
\usepackage{graphicx}
\usepackage{xspace}
\usepackage{xcolor}
\usepackage{stmaryrd}
\usepackage{subcaption}
\usepackage{array}
\usepackage{hyperref}
\usepackage{tablefootnote}
 
\newcommand{\descr}[1]{\vspace{0.1cm}\noindent\textit{#1}}
\usepackage[utf8]{inputenc}
\newtheorem{definition}{Definition}
\newtheorem{theorem}{Theorem}

\usepackage{stmaryrd}
\usepackage[ruled,noline, linesnumbered]{algorithm2e} 
\usepackage{slashbox}
\usepackage{makecell}
\usepackage{xspace}

\usepackage[font=small]{caption} 

\SetAlgoNoLine
\SetAlCapNameFnt{\small}
\SetAlCapFnt{\small}

\newcommand{\mbf}[1]{{\mathbf{#1}}}

\newcommand{\Tcl}{T_{\mathsf{cl}}}
\newcommand{\Tgd}{T_{\mathsf{gd}}}

\newcommand{\TP}{\textit{TP }}
\newcommand{\TN}{\textit{TN }}
\newcommand{\TPR}{\textit{TPR }}
\newcommand{\TNR}{\textit{TNR }}
\newcommand{\Pos}{\textit{P }}
\newcommand{\Neg}{\textit{N }}
\newcommand{\FPR}{\textit{FPR }}

\newcommand{\AUC}{\textit{AUROC }}

\newcommand{\TOPK}{Top-$K$\xspace}

\title{Constrained Differentially Private Federated Learning for Low-bandwidth Devices}
\author[1]{\href{mailto:Raouf Kerkouche <raouf.kerkouche@inria.fr>?Subject=Questions to Raouf}{Raouf~Kerkouche}{}}
\author[2]{\href{mailto:Gergely \'Acs <acs@crysys.hu>?Subject=Questions to Gergely}{Gergely~\'Acs}{}}
\author[1]{\href{mailto:Claude Castelluccia <claude.castelluccia@inria.fr>?Subject=Questions to Claude}{Claude~Castelluccia}{}}
\author[3]{\href{mailto:Pierre Genev\`es <pierre.geneves@cnrs.fr>?Subject=Questions to Pierre}{Pierre~Genev\`es}{}}
\affil[1]{%
    Privatics team, Univ. Grenoble Alpes, Inria, 38000 
}
\affil[2]{%
    Crysys Lab, BME-HIT 
}
\affil[3]{
    Tyrex team, Univ. Grenoble Alpes, CNRS, Inria,
    Grenoble INP, LIG

}

\begin{document}
\maketitle

\begin{abstract}
Federated learning becomes a prominent approach when different entities want to learn collaboratively a common model without sharing their training data. 
However, Federated learning has two main drawbacks. First, it is quite bandwidth inefficient as it involves a lot of message exchanges between the aggregating server and the participating entities. This bandwidth and corresponding processing costs could be prohibitive if the participating entities are, for example, mobile devices.
Furthermore, although federated learning improves privacy by not sharing data, recent attacks have shown that it still leaks information about the training data. 

This paper presents a novel privacy-preserving federated learning scheme. The proposed scheme provides theoretical privacy guarantees, as it is based on Differential Privacy. Furthermore, it optimizes
the model accuracy by constraining the model learning phase on few selected weights. Finally, as shown experimentally, it reduces the upstream \emph{and} downstream bandwidth by up to 99.9\% compared to standard federated learning, making it practical
for mobile systems.

\end{abstract}

\section{Introduction}
\label{sec:intro}

In Machine Learning, different entities may want to collaborate in order to improve their local model accuracy. In traditional machine learning, such collaboration requires to first store all entities' data on a centralized server and then to train a model on it. Such data centralization might be problematic when the data are sensitive and data privacy is required. In order to mitigate this problem, Federated learning, which allows different entities to learn collaboratively a common model without sharing their data, was introduced \citep{ShokriS15,FedAVG}.  Instead of sharing the training data, Federated Learning shares the model parameters between a server, which plays the role of aggregator, and the participating entities. Although Federated Learning improves privacy, model parameters can leak information about the training data. Indeed, \cite{ZhuLH19,idlg,geiping2020inverting} presented some attacks that allow an adversary to reconstruct pieces of the training data of some entities. \cite{NasrSH19} define a membership attack that allows to infer if a particular record is included in the data of a specific entity. Similarly,  \cite{Property_inference} define an attack which aims at inferring if a subgroup of people with a specific property, like for example skin color or ethnicity, is included in the dataset of a particular participating entity.  
A solution to prevent these attacks and provide theoretical guarantees in to use a privacy model called Differential Privacy \citep{Dwork2014book}. 
Differential Privacy has been applied to federated learning in order to protect either each record included in the dataset of any entity (record-level guarantee), or the whole dataset of any entity (client-level guarantee). Unfortunately, it is well-known that Differential Privacy drastically degrades the accuracy of the global model as it requires to add random noise to the gradients (record-level) or to the updates (client-level) of each client. Recent work  by \cite{our_cs} shows that this accuracy penalty can be reduced if the model is compressed, as compression reduces the required amount of noise. Furthermore, \cite{our_cs} show that accuracy can be further improved by adding noise only to the largest update's values as adding noise on values close to 0 is likely to lead to random update values.

Following up on these results, we propose a novel differentially private federated learning solution that improves the model accuracy (1) by updating only a fixed subset of the model weights, and (2) by maintaining the other weights constant. The proposed scheme provides theoretical privacy guarantees, as it is based on Differential Privacy.  Furthermore, it optimizes
the model accuracy by constraining the model learning phase on a few selected weights. As all participants always update the same set of weights and transfer them to the server for aggregation, the proposal can be easily integrated with secure aggregation \citep{BonawitzIKMMPRS16}, which allows parties to add less noise than other decentralized perturbation approaches such as randomized response \citep{ErlingssonPK14} used in local differential privacy.
Moreover, it also reduces the upstream and downstream bandwidth by a factor of 1000 compared to standard federated learning, making it practical
for mobile systems. 
The paper is structured as follows: In Section~\ref{sec:backg} we introduce the necessary background to understand the proposal, in Section~\ref{sec:fl_top_k} we define our solution called FL-TOP and in Section~\ref{sec:fl_top_k_dp} its private extension called FL-TOP-DP.

\section{Background}
\label{sec:backg}
\subsection{Federated Learning (FL-STD)}
\label{FL-STANDARD}

In federated learning \citep{ShokriS15,FedAVG}, multiple parties (clients) build a common machine learning model from union of their training data without sharing them with each other. At each round of the training, a selected set of clients retrieve the global model from the parameter server, update the global model based on their own training data, and send back their updated model to the server. The server aggregates the updated models of all clients to obtain a global model that is re-distributed to some selected parties in the next round.  

In particular, a subset $\mathbb{K}$ of all $N$ clients are randomly selected at each round to update the global model, and $C = |\mathbb{K}| / N$ denotes the fraction of selected clients. At round $t$, a selected client $k \in \mathbb{K}$ executes $\Tgd$ local gradient descent iterations on the common model $\mbf{w}_{t-1}$ using its own training data $D_k$ ($D = \cup_{k\in \mathbb{K}} D_k$), and obtains the updated model $\mbf{w}_{t}^k$, where the number of weights is denoted by $n$ (i.e., $|\mbf{w}_t^{k}| = |\Delta \mbf{w}_t^k| = n$ for all $k$ and $t$). Each client $k$ submits the update $\Delta \mbf{w}_{t}^k = \mbf{w}_{t}^k - \mbf{w}_{t-1}^k$ to the server, which then updates the common model as follows: $\mbf{w}_{t} = \mbf{w}_{t-1} + \sum_{k \in \mathbb{K}} \frac{|D_k|}{\sum_j |D_j|} \Delta \mbf{w}_{t}^k$, where $|D_k|$ is known to the server for all $k$ (a client's update is weighted with the size of its training data).
The server stops training after a fixed number of rounds $\Tcl$, or when the performance of the common model does not improve on a held-out data. 

Note that each $D_k$ may be generated from different distributions (i.e., Non-IID case), that is, any client's local dataset may not be representative of the population distribution \citep{FedAVG}. This can happen, for example, when not all output classes are represented in every client's training data. 
The federated learning of neural networks is summarized in Alg.~\ref{alg:fed_learn}. In the sequel, each client is assumed to use the same model architecture.


The motivation of federated learning is three-fold: first, it aims to provide confidentiality of each participant's training data by sharing only model updates instead of potentially sensitive training data. Second, in order to decrease communication costs, clients can perform multiple local SGD iterations before sending their update back to the server. 
Third, in each round, only a few clients are required to perform local training of the common model, which further diminishes communication costs  and makes the approach especially appealing with large number of clients.

However, several prior works have demonstrated that model updates do leak potentially sensitive information \citep{NasrSH19,Property_inference}. Hence, simply not sharing training data \emph{per se} is not enough to guarantee their confidentiality.

\subsection{Differential Privacy}
\label{sec:DP}
Differential privacy allows a party to privately release information about a dataset:  a function of an input dataset is perturbed, so that any information which can differentiate a record from the rest of the dataset is bounded~\cite{Dwork2014book}.

 \begin{definition}[Privacy loss]
 Let $\mathcal{A}$ be a privacy mechanism which assigns a value $\mathit{Range}(\mathcal{A})$ to a dataset $D$. The privacy loss of $\mathcal{A}$ with datasets $D$ and $D'$ at output $O \in \mathit{Range}(\mathcal{A})$ is a random variable $\mathcal{P}(\mathcal{A},D,D',O) = \log\frac{\Pr[\mathcal{A}(D) = O]}{\Pr[\mathcal{A}(D') = O]}$ 
 where the probability is taken on the randomness of $\mathcal{A}$.
 \label{def:ploss}
 \end{definition}

\begin{definition}[$(\epsilon,\delta)$-Differential Privacy~\citep{Dwork2014book}] 
A privacy mechanism $\mathcal{A}$ guarantees $(\varepsilon, \delta)$-differential privacy if for any database $D$ and $D'$, differing on at most one record, $\Pr_{O \sim \mathcal{A}(D)}[\mathcal{P}(\mathcal{A},D,D',O) > \varepsilon] \leq \delta$. 

\label{def:DP}
\end{definition}

Intuitively, this guarantees that an adversary, provided with the output of $\mathcal{A}$, can draw almost the same conclusions (up to $\varepsilon$ with probability larger than $1 - \delta$) about any record no matter if it is included in the input of $\mathcal{A}$ or not. That is, for any record owner, a privacy breach is unlikely to be due to its participation in the dataset.

\descr{Moments Accountant.} Differential privacy maintains composition; the privacy guarantee of the  $k$-fold adaptive composition  of $\mathcal{A}_{1:k} = \mathcal{A}_1, \ldots, \mathcal{A}_k$ can be computed using the moments accountant method \cite{Abadi}. In particular, it follows from Markov's inequality that $\Pr[\mathcal{P}(\mathcal{A},D,D',O) \geq \varepsilon] \leq \mathbb{E}[\exp(\lambda \mathcal{P}(\mathcal{A},D,D',O))]/\exp(\lambda\varepsilon)$ for any output $O \in \mathit{Range}(\mathcal{A})$ and $\lambda > 0$. 
$\mathcal{A}$ is $(\varepsilon, \delta)$-DP 
with $\delta = \min_{\lambda} \exp(\alpha_{\mathcal{A}}(\lambda) - \lambda \varepsilon)$, where $\alpha_{\mathcal{A}}(\lambda) = \max_{D,D'} \log\mathbb{E}_{O\sim \mathcal{A}(D)}[\exp(\lambda \mathcal{P}(\mathcal{A},D,D',O))]$ is the log of the moment generating function of the privacy loss. The privacy guarantee of the composite mechanism $\mathcal{A}_{1:k}$ can be computed using that $\alpha_{\mathcal{A}_{1:k}}(\lambda) \leq \sum_{i=1}^k \alpha_{\mathcal{A}_{i}}(\lambda)$ \cite{Abadi}. \smallskip 

\descr{Gaussian Mechanism.} 
A fundamental concept of all DP sanitization techniques is the \emph{global sensitivity} of a function~\citep{Dwork2014book}.
\begin{definition}[Global $L_p$-sensitivity] 
For any function $f:\mathcal{D} \rightarrow \mathbb{R}^ n$, the $L_p$-sensitivity of $f$ is
$\Delta_p f = \max_{D, D'} || f(D)-f(D') ||_p$, 
for all $D, D'$ differing in at most one record, where $||\cdot||_p$ denotes the $L_p$-norm.\vspace*{-0.15cm}
\label{def:global_sens}
\end{definition}
\smallskip
The Gaussian Mechanism~\citep{Dwork2014book} 
consists of adding Gaussian noise to the true output of a function.
In particular, for any function $f:\mathcal{D} \rightarrow \mathbb{R}^n$, the Gaussian mechanism is defined as adding i.i.d Gaussian noise with variance $(\Delta_2 f \cdot \sigma)^2$  and zero mean to each coordinate value of  $f(D)$. Recall that the pdf of the Gaussian distribution with mean $\mu$ and variance $\xi^2$ is
$
\mathsf{pdf}_{\mathcal{G}(\mu, \xi)}(x) = \frac{1}{\sqrt{2\pi}\xi} \exp\left(-\frac{(x-\mu)^2}{2 \xi^2}\right) 
$.

In fact, the Gaussian mechanism draws vector values from a multivariate spherical (or isotropic) Gaussian distribution
which is described by random variable $\mathcal{G}(f(D), \Delta_2 f \cdot \sigma\mathbf{I}_n)$, where $n$ is omitted if its unambiguous in the given context.

\section{Federated Pruning}

\label{sec:fl_top_k}

In the standard federated learning scheme (FL-STD, in Section~\ref{sec:backg}), the server sends the latest updated model to a randomly selected set of clients (downstream), and each client sends back its complete model update after local training to the server (upstream) at each round. Knowing that a model has on average millions of parameters (each is a floating point value represented on 32 bits), the network can suffer from large traffic both upstream and downstream. 

Our solution, called FL-TOP, aims to reduce the large amount of network traffic by reducing both downstream and upstream traffic. Moreover, a privacy-preserving extension of this scheme, called FL-TOP-DP, is also proposed, which provides Differential Privacy for the whole training data of every client. 

    


In what follows, we first describe the non-private scheme FL-TOP and then the privacy-preserving FL-TOP-DP. 



\subsection{FL-TOP: Federated Pruning for Compression}





FL-TOP is inspired by the pruning techniques proposed in \cite{dsd} (see Section~\ref{sec:related_work} for more details), and it aims to reduce the amount of parameters exchanged downstream (from the server to the participating entities) and upstream (from the participating entities to the server).  In our scheme, each client updates only a small subset, \TOPK, of the model parameters (weights) at each round. Only the $K$ weight values of these \TOPK parameters are updated during training, and neither the clients nor the server need to transfer the values of the remaining $n-K$ parameters, where $n$ is the total number of parameters. The set of \TOPK parameters do not change over the whole training and are identical for all clients. 
We experimentally show in Section \ref{sec:exp} that, if these $K$ parameters are chosen carefully, the performance penalty is negligible even if $K = 0.005 \cdot n$, that is, $99.5\%$ of the model parameters are pruned. Note that unlike standard pruning techniques, where the set of pruned weights are re-selected after each SGD iteration \citep{dsd}, our scheme always updates the same $K$ parameters.

These \TOPK parameters are selected by the server at the beginning of the protocol. More specifically, the server initializes the model and trains
that with some public data that have a similar distribution as the clients' training data. After a few SGD iterations, the server selects the $K$ parameters which values changed the most.



FL-TOP is described in Alg.~\ref{alg:topk}. First, the server uses public data to identify the set $\mathbb{T}$ of the \TOPK parameters ($K=|\mathbb{T}|$), before starting federated learning. In particular, starting from a public model $\mbf{w}_0$, it accumulates the absolute value of gradients per parameter over $T_{\mathsf{init}}$ SGD iterations, and selects the $K$ parameters with the largest accumulated gradients. 
After that, the values/updates\footnote{weight values for downstream and update/gradients for upstream traffic} of these parameters are the only ones exchanged during the rest of the training between the server and the clients. 

At each round, each selected client $k$ uses the $K$ updated weights $\hat{\mbf{w}}_{t-1}$ received from the server to create a new weight vector $\mbf{w}_{t-1}^{k}$ of size $n$, such that $\mbf{w}_{t-1}^{k}$ is composed from the compressed vector $\hat{\mbf{w}}_{t-1}^{k}$ of size $K \leq n$ (with coordinates in $\mathbb{T}$) and $n-K$ weights from the initialization vector $\mbf{w}_0$. $\mbf{w}_0$ is identical for all participants and can be generated from a shared seed.
Note that when $K=|\mathbb{T}|=n$, the scheme is equivalent to FL-STD. 
The weight vector $\mbf{w}_{t-1}^{k}$ is used to train the client's model. However, only the weights in $\mathbb{T}$ are updated while the remaining ones are kept fixed. To do that, the weights not in $\mathbb{T}$ are reinitialized after each SGD iteration to $\mbf{w}_0$.
The server receives only the values from $\mbf{w}_{t}^{k} - \mbf{w}_{t-1}^{k}$ at coordinates $\mathbb{T}$, denoted by $\mathcal{C}(\mbf{w}_{t}^{k} - \mbf{w}_{t-1}^{k})$ for short, from every client $k$, and updates the common model $\mathbf{w}_t$ with the average of these compressed updates (in Line 12).



\subsection{FL-TOP-DP: Differentially Private Federated Pruning}
\label{sec:fl_top_k_dp}

\subsubsection{Privacy Model}
\label{sec:privacy_model}
We consider an adversary, or a set of colluding adversaries, who can access any update vector sent by the server or any clients at each round of the protocol.  A plausible adversary is a participating entity, i.e. a malicious client or server, that wants to infer the training data used by other participants.
The adversary is \emph{passive} (i.e., honest-but-curious), that is, it follows the learning protocol faithfully. 

Different privacy requirements can be considered depending on what information the adversary aims to infer. In general, private information can be inferred about:
\begin{itemize}
    \item any record (user) in any dataset of any client (\emph{record-level privacy}),
    \item any client/party (\emph{client-level privacy}).
\end{itemize}

To illustrate the above requirements, suppose that several banks build a common model to predict the creditworthiness of their customers. A bank certainly does not want other banks to learn the financial status of any of their customers (record privacy) and perhaps not even the average income of all their customers  (client privacy).

Record-level privacy is a standard requirement used in the privacy literature and is usually weaker than client-level privacy. Indeed, client-level privacy requires to hide any information which is unique to a client including perhaps all its training data.   


We aim at developing a solution that provides \emph{client-level privacy and is also bandwidth efficient}. 
For example, in the scenario of collaborating banks, we aim at protecting any information that is unique to each single bank's training data.
The adversary should not be able to learn from the received model or its updates whether any client's data is involved in the federated run (up to $\varepsilon$ and $\delta$). 
We believe that this adversarial model is reasonable in many practical applications when the confidential information spans over multiple samples in the training data of a single client (e.g., the presence of a group a samples, such as people from a certain race). Differential Privacy guarantees plausible deniability not only to any groups of samples of a client but also to any client in the federated run. Therefore, any negative privacy impact on a party (or its training samples) cannot be attributed to their involvement in the protocol run.

\subsubsection{Operation}
\label{sec:operation}

FL-TOP-DP is described in Alg.~\ref{alg:topk_dp} is very similar to FL-TOP except that each client adds Gaussian noise to its \TOPK model updates to guarantee client-level DP, and applies secure aggregation allowing the server to learn only the aggregated (and noisy) model update. More specifically, each client first calculates its compressed model update $\Delta \mbf{w}_t^k = \mathcal{C}(\mbf{w}_{t}^k- \mbf{w}_{t-1}^k)$ (in Line 25) which is then clipped (in Line 26) to obtain $\Delta \hat{\mbf{w}}_t^k$ with $L_2$-norm at most $S$. After that, random noise $\mbf{z}_{k} \sim \mathcal{G}(0,S\mathbb{\sigma}\mathbf{I}/\sqrt{\mathbb{K}})$ is added to $\Delta \hat{\mbf{w}}_t^k$ such that the sum $\sum_{k \in \mathbb{K}} (\Delta \hat{\mbf{w}}_t^k +  \mbf{z}_{k}) = \sum_{k \in \mathbb{K}} \Delta \hat{\mbf{w}}_t^k + \mathcal{G}(0,S\mathbb{\sigma}\mathbf{I})$ as the sum of Gaussian random variables also follows Gaussian distribution\footnote{More precisely, $\sum_{i}\mathcal{G}(\nu_{i},\xi_{i})=\mathcal{G}(\sum_{i} \nu_{i},\sqrt{\sum_{i}\xi_{i}^{2}})$} and then differential privacy is satisfied where $\varepsilon$ and $\delta$ can be computed using the moments accountant described in Section \ref{sec:DP}. Recall that the \TOPK coordinates in $\mathbb{T}$ are selected and distributed by the server, which is honest-but-curious by assumption.

However, as the noise is inversely proportional to $\sqrt{\mathbb{K}}$, 
$\mbf{z}_{k}$ is likely to be small if $|\mathbb{K}|$ is too large. Therefore, the adversary accessing an individual update $\Delta \hat{\mbf{w}}_t^k + \mbf{z}_{k}$ can almost learn a non-noisy update since $\mbf{z}_{k}$ is small. Hence, each client uses secure aggregation to encrypt its individual update before sending it to the server. Upon reception, the server sums the encrypted updates as:
\begin{align}
\sum_{k\in\mathbb{K}} \mathbf{c}_t^k &= \sum_{k\in\mathbb{K}}\mathsf{Enc}_{K_k}(\Delta \hat{\mbf{w}}_t^k + \mbf{z}_{k}) \nonumber 
 = \sum_{k\in\mathbb{K}} \Delta \hat{\mbf{w}}_t^k + \sum_{k\in\mathbb{K}}\mbf{z}_{k} \nonumber \\
& = \sum_{k\in\mathbb{K}} \Delta \hat{\mbf{w}}_t^k + \mathcal{G}(0,S\mathbb{\sigma}\mathbf{I}) \label{eq:a1}
\end{align}
where $\mathsf{Enc}_{K_k}(\Delta \hat{\mbf{w}}_t^k + \mbf{z}_{k})= \Delta \hat{\mbf{w}}_t^k + \mbf{z}_{k} + \mbf{K}_k \mod p$ and $\sum_{k}\mbf{K}_k=0$ (see \cite{AcsC11,BonawitzIKMMPRS16} for details). Here the modulo is taken element-wise and $p=2^{\lceil \log_{2}(\max_{k}|| \Delta \hat{\mbf{w}}_t^k + \mbf{z}_{k}||_{\infty}|\mathbb{K}|)\rceil}$.
Let $\gamma_t^k = 1/\max\left(1, \frac{||\Delta \mbf{w}_t^k||_2}{S}\right)$. Then, 
\begin{align}
\sum_{k \in \mathbb{K}} \Delta \hat{\mbf{w}}_t^k &= \sum_{k\in \mathbb{K}} \gamma_t^k \Delta \mbf{w}_t^k  \nonumber
 = \sum_{k\in \mathbb{K}} \gamma_t^k \mathcal{C}(\mbf{w}_{t}^k- \mbf{w}_{t-1}^k, \mathbb{T}) \nonumber \\
& = \mathcal{C}(\sum_{k\in \mathbb{K}} \gamma_t^k (\mbf{w}_{t}^k- \mbf{w}_{t-1}^k), \mathbb{T}) \label{eq:a2}
\end{align}
where the last equality comes from the linearity of the compression operation. Indeed, recall that each client selects the values of the \emph{same} \TOPK coordinates from $\mathbb{T}$. 
Plugging Eq.~\eqref{eq:a2} into Eq.~\eqref{eq:a1}. we get that
\begin{align*}
\sum_{k\in\mathbb{K}} \mathbf{c}_t^k = \mathcal{C}(\sum_{k\in \mathbb{K}} \gamma_t^k (\mbf{w}_{t}^k- \mbf{w}_{t-1}^k), \mathbb{T}) + \mathcal{G}(0,S\mathbb{\sigma}\mathbf{I})
\end{align*}

\noindent \textbf{Privacy analysis:}
The server can only access the noisy aggregate  which is sufficiently perturbed to ensure differential privacy; any client-specific information that could be inferred from the noisy aggregate is tracked and quantified by the moments accountant, described in Section~\ref{sec:DP}, as follows. 

Let $\eta_0(x|\xi) =  \mathsf{pdf}_{\mathcal{G}(0, \xi)}(x)$ and $\eta_1(x|\xi) =  (1-C) \mathsf{pdf}_{\mathcal{G}(0, \xi)}(x) + C \mathsf{pdf}_{\mathcal{G}(1, \xi)}(x)$ where $C$ is the sampling probability of a single client in a single round. Let
$
\alpha(\lambda| C) = \log\max(E_1(\lambda, \xi, C), E_2(\lambda, \xi, C)) 
$
where
$
E_1(\lambda,  \xi, C) =  \int_{\mathbb{R}}\eta_0(x|\xi, C) \cdot \left(\frac{\eta_0(x|\xi, C)}{\eta_1(x|\xi, C)}\right)^{\lambda} dx
$ and
$ E_2(\lambda,  \xi, C) = \int_{\mathbb{R}}\eta_1(x|\xi, C) \cdot \left(\frac{\eta_1(x|\xi, C)}{\eta_0(x|\xi, C)}\right)^{\lambda} dx
$.

\begin{theorem}[Privacy of FL-TOP-DP]
\label{thm:dg_privacy}
FL-TOP-DP is $(\min_\lambda  (T_{\mathsf{cl}}\cdot \alpha (\lambda | C)  - \log \delta) /\lambda, \delta)$-DP. 
\end{theorem}
Given a fixed value of $\delta$, $\varepsilon$ is computed numerically  as in \cite{Abadi,MironovTZ19}.

\begin{algorithm}[h]
\small
		\caption{FL-TOP: Federated Learning \label{alg:topk}}
	\DontPrintSemicolon
	{\bf Server:}\;
	\Indp Initialize common model $w_0$\;
	Select set $\mathbb{T}$ of \TOPK updated weights' coordinates via public dataset\;
	\For {$t=1$ \KwTo $\Tcl$}
	{
	    Select $\mathbb{K}$ clients uniformly at random \;
		\For {\textrm{each} client $k$ \textrm{in} $\mathbb{K}$}
		{	
			$\mathbf{c}_t^k = \mathbf{Client}_k(\mathcal{C}(\mbf{w}_{t-1},\mathbb{T}))$\;
		}
		
		$\mbf{w}_{t}=\mbf{w}_{0}$\;
		$j=1$\;
		\For {\textrm{each} coordinate $i$ in $\mathbb{T}$}
		{
		    $\mbf{w}_{t}[i] = \mbf{w}_{t-1}[i] + \sum_{k} \frac{\mathbf{c}_t^k [j]}{|\mathbb{K}|}$\;
		    $j=j+1$\;
	    }
	}
	\KwOut{Global model $\mbf{w}_t$}\;
	\Indm {\bf $\mathbf{Client}_{k}(\hat{\mbf{w}}_{t-1}^k)$:}\;
	\Indp
	
	$\mbf{w}_{t-1}^k=\mbf{w}_{0}$\;
	$j=1$\;
	\For {\textrm{each} coordinate $i$ in $\mathbb{T}$}
	{
	    $\mbf{w}_{t-1}^k[i] = \hat{\mbf{w}}_{t-1}^k[j]$\;
	    $j=j+1$\;
    }
    
	$\mbf{w}_{t}^k = \mathbf{Top_kSGD}(D_k, \mbf{w}_{t-1}^k,\mbf{w}_{0}, \Tgd, \mathbb{T})$\;
	\KwOut{Model update $\mathcal{C}(\mbf{w}_{t}^k- \mbf{w}_{t-1}^k,\mathbb{T})$} 
\end{algorithm}

\begin{algorithm}[h]
\small
	\caption{$\mbf{Top_k}$-Stochastic Gradient Descent \label{alg:sgdk}}
	\DontPrintSemicolon
	\KwIn{$D$ : training data, $\Tgd$ : local epochs, $\mathbf{w}$ : weights, $\mbf{w}_{0}$ : first weights' initialization, $\mathbb{T}$ : set of \TOPK values coordinates .}  
    \For {$t=1$ \KwTo $\Tgd$}
	{
	    Select batch $\mathbb{B}$ from $D$ randomly\;
	    
	    $\mbf{u} =- \eta \nabla f(\mathbb{B}; \mbf{w})$\;
	    
		\For {\textrm{each} coordinate $i$ in $\mathbb{T}$}
		{
		    $\mbf{w}[i] = \mbf{w}[i] + \mbf{u}[i] $\;
	    }
	}
    \KwOut{Model $\mbf{w}$} 
\end{algorithm}

\begin{algorithm}[h]
\small
		\caption{FL-TOP-DP: Federated Learning \label{alg:topk_dp}}
	\DontPrintSemicolon
	{\bf Server:}\;
	\Indp Initialize common model $w_0$\;
	Select set $\mathbb{T}$ of \TOPK updated weights' coordinates via public dataset\;
	\For {$t=1$ \KwTo $\Tcl$}
	{
	    Select $\mathbb{K}$ clients uniformly at random \;
		\For {\textrm{each} client $k$ \textrm{in} $\mathbb{K}$}
		{	
			$\mbf{c}_t^k = \mathbf{Client}_k(\mathcal{C}(\mbf{w}_{t-1},\mathbb{T}))$\;
		}
		
		$\mbf{w}_{t}=\mbf{w}_{0}$\;
		$j=1$\;
		\For {\textrm{each} coordinate $i$ in $\mathbb{T}$}
		{
		    $\mbf{w}_{t}[i] = \mbf{w}_{t-1}[i] + \sum_{k} \frac{ \mbf{c}_{t}^{k}[j]}{|\mathbb{K}|}$\;
		    $j=j+1$\;
	    }
	}
	\KwOut{Global model $\mbf{w}_t$}\;
	\Indm {\bf $\mathbf{Client}_{k}(\hat{\mbf{w}}_{t-1}^k)$:}\;
	\Indp
	
	$\mbf{w}_{t-1}^k=\mbf{w}_{0}$\;
	$j=1$\;
	\For {\textrm{each} coordinate $i$ in $\mathbb{T}$}
	{
	    $\mbf{w}_{t-1}^k[i] = \hat{\mbf{w}}_{t-1}^k[j]$\;
	    $j=j+1$\;
    }
    
	$\mbf{w}_{t}^k = \mathbf{Top_kSGD}(D_k, \mbf{w}_{t-1}^k,\mbf{w}_{0}, \Tgd, \mathbb{T})$\;
	
	$\Delta \mbf{w}_t^k = \mathcal{C}(\mbf{w}_{t}^k- \mbf{w}_{t-1}^k,\mathbb{T})$\;

	$\Delta \hat{\mbf{w}}_t^k = \Delta \mbf{w}_t^k / \max\left(1, \frac{||\Delta \mbf{w}_t^k||_2}{S}\right)$\;
    \KwOut{$\mathsf{Enc}_{K_k}(\mathcal{G}(\Delta \hat{\mbf{w}}_t^k, S \mathbf{I}\sigma /\sqrt{|K|}))$}

\end{algorithm}

\subsubsection{Remarks}

The magnitude of the added Gaussian noise is proportional to the clipping threshold $S$, which is in turn calibrated to the norm of the model update. However, the norm of the model update increases if the model size increases \citep{zhu2020votingbased}, and hence $S$ should be chosen sufficiently large to guarantee fast convergence with large accuracy. On the other hand, too large $S$ also increases the perturbation error caused by the added noise.



\begin{figure*}[h]
  \centering
  \includegraphics[scale=0.35]{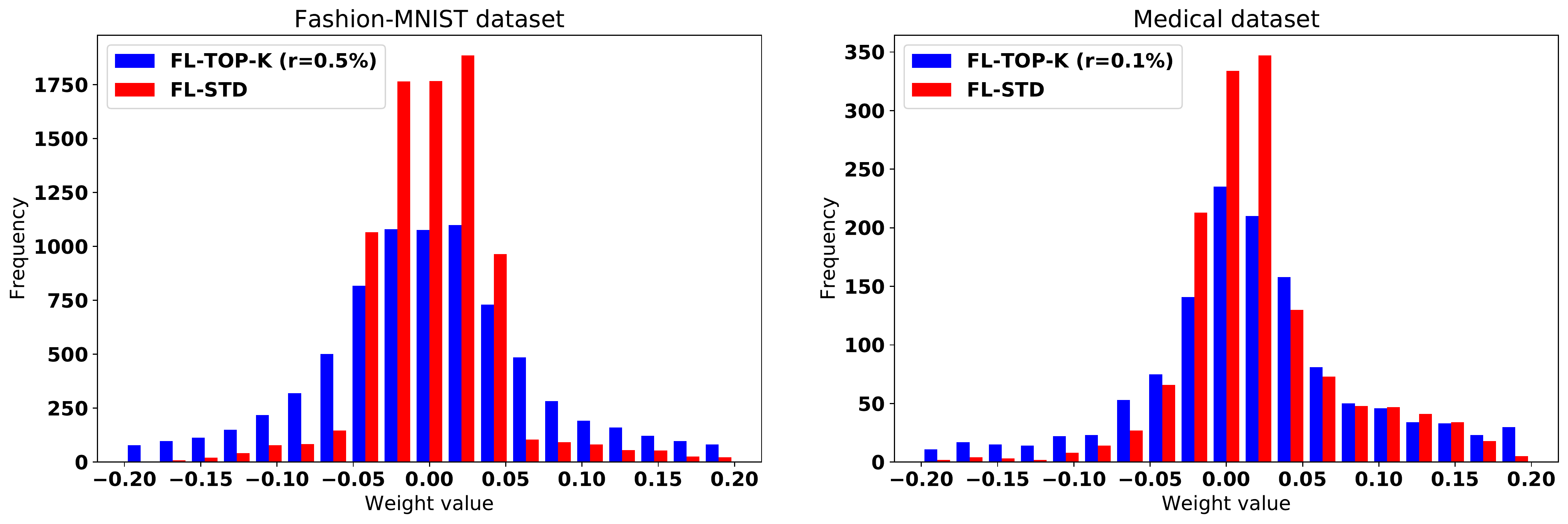}
  \caption{Distributions of the \TOPK weight values (after convergence) for both FL-TOP and FL-STD schemes with the Fashion-MNIST dataset (left) and the medical dataset (right).}
  \label{fig:distribution}
\end{figure*}

FL-TOP aims to diminish this perturbation error by reducing $S$ via compression which also increases the $L_2$-norm of the compressed update vector. This is illustrated in Figure \ref{fig:distribution}, which shows that the norm of the \TOPK coordinates with FL-TOP tend to be larger than with FL-STD (i.e., when all coordinates get updated not only the \TOPK). Therefore, besides decreasing the magnitude of the added noise, FL-TOP also decreases the relative error on the retained parameters. These together decrease the perturbation error caused by the added noise. 



Notice that there exist other alternatives to identify the \TOPK coordinates in a privacy-preserving manner than using a public dataset. For example, every client can select the \TOPK parameters with the largest magnitude during the first rounds locally, and send them to the server for aggregation. More specifically,
each client creates a parameter vector with size $n$, where the \TOPK coordinates are set to 1 while the rest are kept 0. Then, these binary vectors are noised and aggregated by the server like in Section \ref{sec:operation}. In the rest of the training, all participants exchange only the updates and weights of the these \TOPK parameters like in FL-TOP.
However, aside from consuming more privacy budget, this approach also has lower accuracy than our proposal according to our tests. Moreover, it has larger communication cost in the initialization phase when the \TOPK parameters are identified and the whole binarized parameter vector is sent for aggregation.

\section{Experimental Results}
\label{sec:exp}

The goal of this section is to evaluate the performance of our proposed schemes FL-TOP and FL-TOP-DP on a benchmark dataset and a realistic in-hospital mortality prediction scenario. We aim at evaluating their performance with different levels of compression and comparing them with the performance of the following learning protocols\footnote{More baselines are considered but due to the lack of space, we have decided to present only those which return the best results. All other results can be found in the appendix( Section~\ref{sec:furthmore_experiments}).}:
\begin{itemize}
    \item FL-STD: The Standard Federated Learning scheme as described in Section \ref{FL-STANDARD} (see Alg.~\ref{alg:fed_learn}). 
    \item FL-BASIC: A Federated Learning scheme that updates a random subset of parameters instead of the \TOPK parameters at each SGD iteration. This subset is re-selected at the beginning of each new round.
    The $n-k$ non-selected parameters are still reinitialized after each SGD update as in FL-TOP. 
    \item FL-CS: A Federated Learning scheme that uses Compressive sensing (CS) to compress model updates from \cite{our_cs}. See Section \ref{sec:related_work} for more details.
\end{itemize}

Note that all compression operators in the baselines are linear (just like FL-TOP-DP), and hence they can also be used with secure aggregation. Similarly to FL-TOP-DP, the private extensions (i.e., FL-BASIC-DP and FL-CS-DP) also clip and then noise the compressed updates. 

We evaluate the above learning algorithms on the well-known Fashion-MNIST dataset \citep{Fashion-MNIST} and on the Premier Healthcare Database, which is  a real-world medical dataset of 1.2 millions of US hospital patients \footnote{\href{https://www.premierinc.com/newsroom/education/premier-healthcare-database-whitepaper}{https://www.premierinc.com/newsroom/education/premier-healthcare-database-whitepaper}}. More details can be found in Appendix~\ref{sec:medical_dataset_desc} and Appendix~\ref{sec:fashion_mnist_desc}.

Recall that the \TOPK weights are selected before starting the federated learning process using public data. For Fashion-MNIST, we randomly select  a batch with size 10 from MNIST dataset \citep{MNIST} described in Appendix~\ref{sec:mnist_desc}. For the medical dataset, we did not find any public dataset with the same features as ours, and for this reason, we selected randomly from the dataset a batch of 356 patients\footnote{Reduced to 24 patients when we train via downsampling with 12 patients for each class}. This set is used only by the server and never by any client. Afterwards, the server performs $T_{\mathsf{init}}$ SGD iterations starting from the model parameters $\mathbf{w}_0$ on the same batch to identify the \TOPK weights.
We experimentally show later that even these small batches are enough for the server to find a good set of \TOPK weights. 



In order to select the clipping threshold $S$, 
the server executes a single training round locally, which is composed of $\Tgd$ SGD iterations starting from the model parameters $\mathbf{w}_0$, using the batch from the public data. The clipping threshold $S$ is set to the $L_2$-norm of the \TOPK weight update obtained for this single training round. For FL-BASIC-DP, the same steps are repeated for 100 times, where a new random set of trainable weights with size $K$ are selected each time, which yields 100 $L_2$-norm values. $S$ is set to the median of these $L_2$-norm values. We think that this approach is more fair, because the set of trainable weights is re-selected at each round in FL-BASIC-DP. The computed values of $S$ can be found in Table~\ref{tab:Sensitivity_fashion_mnist} and Table~\ref{tab:Sensitivity_medical_data} for Fashion-MNIST and Medical dataset, respectively.  
More information about the model architectures and the hyper-parameter selection can be found in Appendix \ref{sec:app}.

\begin{table*}[!ht]
    \begin{minipage}{.5\linewidth}
    \centering
    \scalebox{0.6}{
    \begin{tabular}{|c|c|c|c|c|c|c|}
        \hline
         \multirow{4}{*}{\emph{$r$}} & \multirow{4}{*}{Algorithms} & \multicolumn{5}{c|}{\emph{Performance}} \\
         \cline{3-7}
         & &  \multirow{3}{*}{\emph{Accuracy}}  & \multirow{3}{*}{\emph{Round}} & \emph{Downstream} & \emph{Upstream} & \multirow{3}{*}{\emph{$\epsilon$}} \\
         
         & &                &                 &           \emph{Cost}                &     \emph{Cost}              &                     \\
         
         & &               &                   &            (Kilobyte)           &        (Kilobyte)             &                       \\
         
         \hline

        \multirow{8}{*}{$0.5\%$} &  FL-BASIC   &  0.65  & 193 & 21402.03 & 107 & N/A  \\
        \cline{2-7}
        &  FL-CS  &  0.57  & 185 & 20514.9 & 102.56 & N/A  \\
        \cline{2-7}
        &  \textbf{FL-TOP}   & \textbf{0.82}  & 200 & \textbf{110.88} & \textbf{110.88} & N/A \\
        \cline{2-7}
        \cline{2-7}
        &  FL-BASIC-DP   &  0.59  & 200 & 22178.27 & 110.88 & 1\\
        \cline{2-7}
        &  FL-CS-DP  &  0.53  & 200 & 22178.27 & 110.88 & 1\\
        \cline{2-7}
        &  \textbf{FL-TOP-DP}   & \textbf{0.81} & 200 & \textbf{110.88}  & \textbf{110.88} & \textbf{1}\\
        \hline 
        \hline
        \multirow{8}{*}{$5\%$} &  FL-BASIC   &  0.78 & 196 & 21734.70 & 1086.73 & N/A  \\
        \cline{2-7}
        &  FL-CS  &  0.82  & 200 & 22178.27 & 1108.91 & N/A  \\
        \cline{2-7}
        &  \textbf{FL-TOP}   & \textbf{0.84}  & 200 & \textbf{1108.91} & \textbf{1108.91} & N/A \\
        \cline{2-7}
        \cline{2-7}
        &  FL-BASIC-DP   &  0.76  & 195 & 21623.81 & 1081.18 & 0.99 \\
        \cline{2-7}
        &  FL-CS-DP  &  0.78  & 160 & 17742.61 & 887.13 & 0.94 \\
        \cline{2-7}
        &  \textbf{FL-TOP-DP}   &  \textbf{0.81} & 152 & \textbf{842.77}  & \textbf{842.77} & \textbf{0.92}\\
        \hline 
        \hline
        \multirow{8}{*}{$10\%$} &  FL-BASIC   &  0.81  & 196 & 21734.70 & 2173.47 &N/A  \\
        \cline{2-7}
        &  FL-CS  &  0.85  & 182 & 20182.22 & 2018.22 & N/A  \\
        \cline{2-7}
        &  \textbf{FL-TOP}   & \textbf{0.85}  & 199 & \textbf{2206.74} & \textbf{2206.74} &N/A \\
        \cline{2-7}
        \cline{2-7}
        &  FL-BASIC-DP   &  0.79  & 189 & 20958.46 & 2095.85 & 0.98\\
        \cline{2-7}
        &  FL-CS-DP  &  0.72  & 167 & 18518.85 & 1851.89 & 0.95\\
        \cline{2-7}
        &  \textbf{FL-TOP-DP}   &  \textbf{0.80} & 157 & \textbf{1740.99} & \textbf{1740.99} & \textbf{0.93}\\
        \hline 
        \hline
        \multirow{2}{*}{$100\%$} &  FL-STD  &  0.86  & 200 & 22178.27 & 22178.27 & N/A  \\
        \cline{2-7}
        &  FL-STD-DP  &  0.56 & 60 & 6653.48  & 6653.48 & 0.76 \\
        \hline
        
    \end{tabular}}
    \caption{Summary of results on Fashion-MNIST dataset.}
    \label{tab:description_results_Fashion_MNIST_reduced}
\vspace{-.3cm}
    \end{minipage}\hfill
    \begin{minipage}{.5\linewidth}
    \centering
    \scalebox{0.6}{
    \begin{tabular}{|c|c|c|c|c|c|c|c|}
        \hline
         \multirow{4}{*}{\emph{$r$}} & \multirow{4}{*}{Algorithms} & \multicolumn{6}{c|}{\emph{Performance}} \\
         \cline{3-8}
         & &   \multirow{3}{*}{\emph{Bal\_Acc}} & \multirow{3}{*}{\emph{AUROC}}  & \multirow{3}{*}{\emph{Round}} & \emph{Downstream} & \emph{Upstream} & \multirow{3}{*}{\emph{$\epsilon$}} \\
         & &                   &               &              &  \emph{Cost}                    &     \emph{Cost}                &                   \\
         &&                    &               &              &  (Kilobyte)                     &      (Kilobyte)                &     \\
         \hline

        \multirow{8}{*}{$0.1\%$} &  FL-BASIC  & 0.51 & 0.51 & 99 & 11829.42 & 11.82 & N/A \\
        \cline{2-8}
        &  FL-CS &  0.53  & 0.55 & 100 & 11948.91 & 11.94 &N/A \\
        \cline{2-8}
        &  \textbf{FL-TOP} & \textbf{0.69}  & \textbf{0.76} & 68 & \textbf{8.12} & \textbf{8.12} &N/A \\
        \cline{2-8}
        \cline{2-8}
        &  FL-BASIC-DP  & 0.50 & 0.49 & 100 & 11948.91 & 11.94 & 1  \\
        \cline{2-8}
        &  FL-CS-DP &  0.51  & 0.51 & 99 & 11829.42 & 11.82 & 1\\
        \cline{2-8}
        &  \textbf{FL-TOP-DP} &  \textbf{0.69}  & \textbf{0.76} & 85 & \textbf{10.15} & \textbf{10.15} & \textbf{0.97} \\
        \hline 
        \hline
        \multirow{8}{*}{$5\%$} &  FL-BASIC  & 0.72 & 0.80 & 100 & 11948.91 & 597.45 &N/A\\
        \cline{2-8}
        &  FL-CS &   0.73 & 0.81 & 98 & 11709.93 & 585.5 & N/A \\
        \cline{2-8}
        &  \textbf{FL-TOP} &  \textbf{0.72}  & \textbf{0.80} & 95 & \textbf{567.57} & \textbf{567.57} &N/A \\
        \cline{2-8}
        \cline{2-8}
        &  FL-BASIC-DP  & 0.69 & 0.76 & 100 & 11948.91 & 597.45 & 1 \\
        \cline{2-8}
        &  FL-CS-DP &  0.69  & 0.76 & 100 & 11948.91 & 597.45 & 1\\
        \cline{2-8}
        &  \textbf{FL-TOP-DP} & \textbf{0.68} & \textbf{0.75} & 23 & \textbf{137.41} & \textbf{137.41} & \textbf{0.79}\\
        \hline 
        \hline
        
        \multirow{8}{*}{$10\%$} &  FL-BASIC  & 0.74 & 0.81 & 100 & 11948.91 & 1194.89 &N/A \\
        \cline{2-8}
        &  FL-CS &  0.74  & 0.82 & 100 & 11948.91 & 1194.89 & N/A \\
        \cline{2-8}
        &  \textbf{FL-TOP} & \textbf{0.74}  & \textbf{0.82} & 90 & \textbf{1075.40} & \textbf{1075.40} &N/A \\
        \cline{2-8}
        \cline{2-8}
        &  FL-BASIC-DP  & 0.69 & 0.76 & 99 & 11829.42 & 1182.94 & 1 \\
        \cline{2-8}
        &  FL-CS-DP &  0.69  & 0.76 & 96 & 11470.95 & 1147.09 & 0.99\\
        \cline{2-8}
        &  \textbf{FL-TOP-DP} &  \textbf{0.68}  & \textbf{0.74} & 23 & \textbf{274.82} & \textbf{274.82} & \textbf{0.79}\\
        \hline 
        \hline

        \multirow{2}{*}{$100\%$} &  FL-STD  & 0.74 & 0.82 & 99 & 11829.42 & 11829.42 &N/A \\
        \cline{2-8}
        &  FL-STD-DP   & 0.66 & 0.72 & 62 & 7408.32 & 7408.32 & 0.91 \\
        \hline
        
    \end{tabular}}
    \caption{Summary of results on Medical dataset.}
    \label{tab:description_results_Medical_data_reduced}
\vspace{-.3cm}
\end{minipage} 
\end{table*}

\subsection{Results}
\label{sec:results}

Figure~\ref{fig:distribution} displays the distribution of the \TOPK updated weights for FL-TOP and FL-STD at the end of the training. We select the weights when each scheme reached the best accuracy over 200 and best balanced accuracy\footnote{See Appendix~\ref{sec:metrics} for more details.} over 100 rounds for fashion-MNIST and the medical dataset, respectively. We choose the smallest compression ratio $r$ that leads to the best accuracy for the FL-TOP-DP scheme. Table~\ref{tab:description_results_Fashion_MNIST_reduced} shows that FL-TOP-DP reaches the best accuracy, 0.81, when $r=0.5$\% on fashion-MNIST and reaches the best accuracy, 0.69,  when $r=0.1$\% on the medical dataset. Both figures validate the intuition that by constraining the model to update only a small set $K$ of the total weights, these \TOPK become more important and reach larger values. This result is important when differential privacy is used as it leads to larger value-to-noise level and therefore better performance.

Table~\ref{tab:description_results_Fashion_MNIST_reduced} represents the best accuracy over 200 rounds for each scheme on the Fashion-MNIST dataset. $\mathit{Round}$ corresponds to the round when the best accuracy is reached and $\mathit{Cost}$ is the average bandwidth consumption calculated as: $r \times n \times 32 \times \mathit{Round} \times C$, where 32 is the number of bits necessary to represent a float value, $n$ is the uncompressed model size, $r=\frac{|\mathbb{T}|}{n}$, $|\mathbb{T}|$ is the compressed model size, $C$ is the sampling probability of a client, and $\mathit{Round}$ is the round when we get the the best accuracy.

Table~\ref{tab:description_results_Medical_data_reduced} represents the best balanced accuracy over 100 rounds for each scheme on the Medical dataset. $AUROC$ (area under the receiver operating characteristic curve - see Appendix~\ref{sec:metrics}) corresponds to the $AUROC$ value when the best balanced accuracy is reached.

These figures show that the proposed non-private scheme FL-TOP has similar accuracy than the standard scheme FL-STD but reduces the bandwidth cost significantly.
For example, with the Fashion-MNIST dataset, the FL-TOP accuracy reaches 0.85  when the compression ratio $r=10\%$. In comparison, the standard FL-STD scheme reaches an accuracy of 0.86\% but consumes 10 times more bandwidth. Furthermore, although FL-CS reaches the same accuracy than FL-TOP and consumes slightly less bandwidth upstream (9\% less), its required downstream bandwidth is about 10 times larger (See Table~\ref{tab:description_results_Fashion_MNIST_reduced} for more details).
The results on the medical dataset are quite similar. In fact, FL-TOP achieves its best balanced accuracy (0.74) and AUROC (0.82) when $r=10\%$ while the FL-STD scheme obtains similar performance but required about 11 times more upsteam and downstream bandwidth cost. FL-CS achieves similarly accuracy at $r=10\%$ as FL-TOP but its downstream required bandwidth is about 11 times larger (see Table~\ref{tab:description_results_Medical_data_reduced} for more details).


The results also show that not only our privacy-preserving solution FL-TOP-DP provides strong privacy guarantee (with $\epsilon$ values smaller than 1) but that it outperforms the other schemes in term of accuracy and bandwidth, for both datasets. 
For example, with Fashion-MNIST, our scheme achieves an accuracy of $0.81$ when  $r=0.5\%$ while the baseline scheme, FL-BASIC-DP, achieves an accuracy of $0.79$ when $r=10\%$ and requires 189 times more downstream bandwidth and 18 times more upstream bandwidth.  With the medical dataset,  FL-TOP-DP reaches the best balanced accuracy 0.69 and best AUROC $0.76$ for a compression ratio of  $r=0.1\%$ while FL-BASIC-DP an FL-CS-DP achieves the same performance at $r=5\%$.  Note that FL-STD-DP performs very poorly as noise has to be added to the all weights of the model and the sensitivity is large (see Table~\ref{tab:description_results_Medical_data_reduced}).

\section{Related work}
\label{sec:related_work}

\noindent \textbf{Privacy of Federated Learning:}
The concept of Client-based Differential Privacy has been introduced in \cite{Client-DP-McMahan} and \cite{Client-DP-ETH-Zurich}, where the goal is to hide any information that is specific to a single client's training data. These algorithms noise the contribution of a single client instead of a single record in the client's dataset. The noise is added by the server, hence, unlike our solution, these works assume that the server is trusted. 

Recently, \cite{Liu_2020} also proposed to add noise only to the update of the \TOPK model parameters a la local-DP. In local-DP, each client adds larger noise that what is necessary to guarantee DP for the \emph{aggregated} model update without using secure aggregation. Therefore, the common model is less accurate than with our scheme. In addition,  \cite{Liu_2020} uses two epsilon budgets; one for selecting \TOPK parameters per client, and the second for perturbing these selected \TOPK parameters. By contrast, we select the \TOPK parameters via public data without sacrificing any privacy budget. Finally, their solution is also less bandwidth efficient than ours: as the \TOPK parameters differ for each client and at each round, the client cannot send only the \TOPK parameters values because the server will not be able to identify which value corresponds to which \TOPK parameter. For this reason, the client has to send a sparse vector with only \TOPK perturbed values and all remaining parameters set to 0. Therefore, the quantization of the non-\TOPK parameters is performed only during the upstream (from client to server) without compressing any downstream traffic. As opposed to this, in our solution, only the weights/updates of the \TOPK parameters are transferred downstream/upstream.

Recently, \cite{our_cs} proposed to use Compressive sensing (CS) in the context of federated learning in order to compress model updates meanwhile providing  client-level DP. Assuming that the model update is already sparse in the time domain, the noise is added to its largest Fourier coefficients in a distributed manner, and the noisy aggregate is reconstructed with standard optimization techniques. 
Likewise our solution, this work also uses secure aggregation by exploiting the linearity of CS. However, the reconstruction process can be slow for large models, and therefore our solution is more scalable. Moreover, it can only compress the upstream traffic.

\smallskip

\noindent \textbf{Bandwidth Optimization in Federated Learning:}
Different quantization methods have been proposed to save the bandwidth and reduce the communication costs in federated learning. They can be divided into two main groups: unbiased and biased methods. The unbiased approximation techniques use probabilistic quantization schemes to compress the stochastic gradient and attempt to approximate the true gradient value as much as possible \citep{QSGD, TERNGRAD, ATOMO, Quant_Fed}. 
However, biased approximations of the stochastic gradient can still guarantee convergence both in theory and practice \citep{SIGNSGD, LinHM0D18, SeideFDLY14}. 
SignSGD \cite{SIGNSGD} a quantization protocol allows to compress during downstream and upstream traffic but requires the use of all the clients at each round which is not realistic in the context of federated settings because each client is available only during few rounds \cite{kairouz2019advances}.

A different line of works exploit the sparsity of model updates to compress them.
 \cite{mohamedamiri,mohamedamiri2}  proposed to use a compressive sensing for federated learning in order to compress model updates without privacy guarantees. However, they assume that all clients participate in each round (as they maintain an error accumulation vector at each client due to the compression scheme), but as discussed in \cite{kairouz2019advances} this assumption is not always realistic. 
Sketching was adapted to federated learning for the purpose of compressing model updates in \cite{ivkin2019communication} and \cite{rothchild2020fetchsgd}. The authors proposed to use Count-Sketch from \cite{charikar2002finding} to retrieve the largest weights in the update vector on the server side. 
However, it is unclear how these works can be extended with privacy guarantees. Moreover, unlike our technique, they do not compress downstream traffic.

Constraining the weights to have a specific distribution has already been studied. In \cite{dsd}, for example, the authors use pruning techniques to create a sparse model at the end of the training. After each SGD iteration, the authors zero-out all the weights with an absolute value smaller than a threshold. Iterating the process leads to a sparse model with only some absolute weight values larger than 0. Similarly, \cite{Joshua_binary} aim to create a model with binary weights such that at the end of the training all the weights are close to $1$ or $-1$. After each SGD update, the authors take the sign of the weights before the next update. After some iterations, the weight values become close to the interval limits $-1$ and $1$.

In \cite{lottery_ticket}, a new hypothesis claims that there exists a sub-network which, if trained separately, can achieve similar performance as the complete network model which contains that. To find such a sub-network, one has to follow a simple iterative procedure: train the complete network, prune the smallest weights, and then reinitialize the remaining weights to their original values. These steps are repeated iteratively. This approach was extended to federated learning in \cite{li2020lotteryfl}.



\section{Conclusion}

This paper presents a novel privacy-preserving federated learning scheme that reduces bandwidth, latency and
therefore power consumption.
The proposed scheme is based on Differential Privacy and therefore provides theoretical privacy guarantees. Furthermore, it optimizes
the model accuracy by constraining the model learning phase on few selected weights. We show experimentally, using a public 
dataset called Fashion-MNIST and a real world medical dataset of 1.2 millions of US hospital patients, that it reduces the 
upstream and downstream bandwidth by up to 99.9\% compared to standard federated learning, making it practical
for constrained and mobile devices.






\clearpage

\bibliography{references}

\begin{thebibliography}{52}
\providecommand{\natexlab}[1]{#1}
\providecommand{\url}[1]{\texttt{#1}}
\expandafter\ifx\csname urlstyle\endcsname\relax
  \providecommand{\doi}[1]{doi: #1}\else
  \providecommand{\doi}{doi: \begingroup \urlstyle{rm}\Url}\fi

\bibitem[Abadi et~al.(2015)Abadi, , et~al.]{TensorFlow}
Mart\'{\i}n Abadi, , et~al.
\newblock {TensorFlow}: Large-scale machine learning on heterogeneous systems,
  2015.
\newblock Software available from tensorflow.org.

\bibitem[Abadi et~al.(2016)Abadi, Chu, Goodfellow, McMahan, Mironov, Talwar,
  and Zhang]{Abadi}
Martin Abadi, Andy Chu, Ian Goodfellow, H.~Brendan McMahan, Ilya Mironov, Kunal
  Talwar, and Li~Zhang.
\newblock Deep learning with differential privacy.
\newblock In \emph{ACM CCS}, 2016.

\bibitem[{\'{A}}cs and Castelluccia(2011)]{AcsC11}
Gergely {\'{A}}cs and Claude Castelluccia.
\newblock I have a dream! (differentially private smart metering).
\newblock In \emph{{IH}}, 2011.

\bibitem[Akosa(2017)]{akosa2017predictive}
Josephine Akosa.
\newblock Predictive accuracy: a misleading performance measure for highly
  imbalanced data.
\newblock In \emph{Proceedings of the SAS Global Forum}, pages 2--5, 2017.

\bibitem[Alistarh et~al.(2016)Alistarh, Li, Tomioka, and Vojnovic]{QSGD}
Dan Alistarh, Jerry Li, Ryota Tomioka, and Milan Vojnovic.
\newblock {QSGD:} randomized quantization for communication-optimal stochastic
  gradient descent.
\newblock 2016.

\bibitem[Amiri and G{\"{u}}nd{\"{u}}z(2019{\natexlab{a}})]{mohamedamiri}
Mohammad~Mohammadi Amiri and Deniz G{\"{u}}nd{\"{u}}z.
\newblock Machine learning at the wireless edge: Distributed stochastic
  gradient descent over-the-air.
\newblock 2019{\natexlab{a}}.

\bibitem[Amiri and G{\"{u}}nd{\"{u}}z(2019{\natexlab{b}})]{mohamedamiri2}
Mohammad~Mohammadi Amiri and Deniz G{\"{u}}nd{\"{u}}z.
\newblock Federated learning over wireless fading channels.
\newblock 2019{\natexlab{b}}.

\bibitem[Avati et~al.(2018)Avati, Jung, Harman, Downing, Ng, and
  Shah]{Avati2018}
Anand Avati, Kenneth Jung, Stephanie Harman, Lance Downing, Andrew Ng, and
  Nigam~H. Shah.
\newblock Improving palliative care with deep learning.
\newblock \emph{BMC Medical Informatics and Decision Making}, 2018.

\bibitem[Bekkar et~al.(2013)Bekkar, Djema, and Alitouche]{Balanced_acc_2}
Mohamed Bekkar, Hassiba Djema, and T.A. Alitouche.
\newblock Evaluation measures for models assessment over imbalanced data sets.
\newblock \emph{Journal of Information Engineering and Applications},
  3:\penalty0 27--38, 01 2013.

\bibitem[Bernstein et~al.(2018)Bernstein, Wang, Azizzadenesheli, and
  Anandkumar]{SIGNSGD}
Jeremy Bernstein, Yu{-}Xiang Wang, Kamyar Azizzadenesheli, and Anima
  Anandkumar.
\newblock signsgd: compressed optimisation for non-convex problems.
\newblock 2018.

\bibitem[Bonawitz et~al.(2016)]{BonawitzIKMMPRS16}
Keith Bonawitz et~al.
\newblock Practical secure aggregation for federated learning on user-held
  data.
\newblock 2016.

\bibitem[Brodersen et~al.(2010)Brodersen, Ong, Stephan, and
  Buhmann]{Balanced_acc_1}
Kay~Henning Brodersen, Cheng~Soon Ong, Klaas~Enno Stephan, and Joachim~M
  Buhmann.
\newblock The balanced accuracy and its posterior distribution.
\newblock In \emph{20th International Conference on Pattern Recognition}. IEEE,
  2010.

\bibitem[Charikar et~al.(2002)Charikar, Chen, and
  Farach-Colton]{charikar2002finding}
Moses Charikar, Kevin Chen, and Martin Farach-Colton.
\newblock Finding frequent items in data streams.
\newblock In \emph{International Colloquium on Automata, Languages, and
  Programming}, pages 693--703. Springer, 2002.

\bibitem[Chollet et~al.(2015{\natexlab{a}})]{KERAS}
Fran\c{c}ois Chollet et~al.
\newblock Keras.
\newblock \url{https://keras.io}, 2015{\natexlab{a}}.

\bibitem[Chollet et~al.(2015{\natexlab{b}})]{KERAS_datasets}
Fran\c{c}ois Chollet et~al.
\newblock Keras datasets.
\newblock \url{https://keras.io/datasets/}, 2015{\natexlab{b}}.

\bibitem[Courbariaux et~al.(2016)Courbariaux, Hubara, Soudry, El-Yaniv, and
  Bengio]{Joshua_binary}
Matthieu Courbariaux, Itay Hubara, Daniel Soudry, Ran El-Yaniv, and Yoshua
  Bengio.
\newblock Binarized neural networks: Training deep neural networks with weights
  and activations constrained to +1 or -1, 2016.

\bibitem[CUADRADO(2019)]{ICD9}
Marta~TERRON CUADRADO.
\newblock Icd-9-cm: International classification of diseases, ninth revision,
  clinical modification.
\newblock
  \url{https://ec.europa.eu/cefdigital/wiki/display/EHSEMANTIC/ICD-9-CM\%3A+International+Classification+of+Diseases\%2C+Ninth+Revision\%2C+Clinical+Modification},
  2019.

\bibitem[Dwork and Roth(2014)]{Dwork2014book}
Cynthia Dwork and Aaron Roth.
\newblock {The Algorithmic Foundations of Differential Privacy}.
\newblock \emph{Foundations and Trends in Theoretical Computer Science},
  9\penalty0 (3--4), 2014.

\bibitem[Erlingsson et~al.(2014)Erlingsson, Pihur, and
  Korolova]{ErlingssonPK14}
{\'{U}}lfar Erlingsson, Vasyl Pihur, and Aleksandra Korolova.
\newblock {RAPPOR:} randomized aggregatable privacy-preserving ordinal
  response.
\newblock In Gail{-}Joon Ahn, Moti Yung, and Ninghui Li, editors,
  \emph{Proceedings of the 2014 {ACM} {SIGSAC} Conference on Computer and
  Communications Security, Scottsdale, AZ, USA, November 3-7, 2014}, pages
  1054--1067. {ACM}, 2014.
\newblock \doi{10.1145/2660267.2660348}.
\newblock URL \url{https://doi.org/10.1145/2660267.2660348}.

\bibitem[{Fejza} et~al.(2018){Fejza}, {Genevès}, {Layaïda}, and
  {Bosson}]{Papier_Amela}
A.~{Fejza}, P.~{Genevès}, N.~{Layaïda}, and J.~{Bosson}.
\newblock Scalable and interpretable predictive models for electronic health
  records.
\newblock In \emph{{DSAA}}, 2018.

\bibitem[Frankle and Carbin(2018)]{lottery_ticket}
Jonathan Frankle and Michael Carbin.
\newblock The lottery ticket hypothesis: Training pruned neural networks.
\newblock 2018.

\bibitem[Geiping et~al.(2020)Geiping, Bauermeister, Dröge, and
  Moeller]{geiping2020inverting}
Jonas Geiping, Hartmut Bauermeister, Hannah Dröge, and Michael Moeller.
\newblock Inverting gradients -- how easy is it to break privacy in federated
  learning?, 2020.

\bibitem[Geyer et~al.(2017)Geyer, Klein, and Nabi]{Client-DP-ETH-Zurich}
Robin~C. Geyer, Tassilo Klein, and Moin Nabi.
\newblock Differentially private federated learning: {A} client level
  perspective.
\newblock 2017.

\bibitem[Han et~al.(2016)Han, Pool, Narang, Mao, Gong, Tang, Elsen, Vajda,
  Paluri, Tran, Catanzaro, and Dally]{dsd}
Song Han, Jeff Pool, Sharan Narang, Huizi Mao, Enhao Gong, Shijian Tang, Erich
  Elsen, Peter Vajda, Manohar Paluri, John Tran, Bryan Catanzaro, and
  William~J. Dally.
\newblock Dsd: Dense-sparse-dense training for deep neural networks, 2016.

\bibitem[He and Garcia(2009)]{he2009learning}
Haibo He and Edwardo~A Garcia.
\newblock Learning from imbalanced data.
\newblock \emph{IEEE Transactions on knowledge and data engineering}, 2009.

\bibitem[Ivkin et~al.(2019)Ivkin, Rothchild, Ullah, Stoica, Arora,
  et~al.]{ivkin2019communication}
Nikita Ivkin, Daniel Rothchild, Enayat Ullah, Ion Stoica, Raman Arora, et~al.
\newblock Communication-efficient distributed sgd with sketching.
\newblock In \emph{Advances in Neural Information Processing Systems}, pages
  13144--13154, 2019.

\bibitem[Kairouz et~al.(2019)]{kairouz2019advances}
Peter Kairouz et~al.
\newblock Advances and open problems in federated learning.
\newblock 2019.

\bibitem[Kerkouche et~al.(2020)Kerkouche, Ács, Castelluccia, and
  Genevès]{our_cs}
Raouf Kerkouche, Gergely Ács, Claude Castelluccia, and Pierre Genevès.
\newblock Compression boosts differentially private federated learning, 2020.
\newblock To appear in EuroS\&P 2021.

\bibitem[Konecn{\'{y}} et~al.(2016)Konecn{\'{y}}, McMahan, Yu, Richt{\'{a}}rik,
  Suresh, and Bacon]{Quant_Fed}
Jakub Konecn{\'{y}}, H.~Brendan McMahan, Felix~X. Yu, Peter Richt{\'{a}}rik,
  Ananda~Theertha Suresh, and Dave Bacon.
\newblock Federated learning: Strategies for improving communication
  efficiency.
\newblock 2016.

\bibitem[LeCun and Cortes(2010)]{MNIST}
Yann LeCun and Corinna Cortes.
\newblock {MNIST} handwritten digit database.
\newblock 2010.
\newblock URL \url{http://yann.lecun.com/exdb/mnist/}.

\bibitem[Li et~al.(2020)Li, Sun, Wang, Duan, Li, Chen, and Li]{li2020lotteryfl}
Ang Li, Jingwei Sun, Binghui Wang, Lin Duan, Sicheng Li, Yiran Chen, and Hai
  Li.
\newblock Lotteryfl: Personalized and communication-efficient federated
  learning with lottery ticket hypothesis on non-iid datasets, 2020.

\bibitem[Lin et~al.(2018)Lin, Han, Mao, Wang, and Dally]{LinHM0D18}
Yujun Lin, Song Han, Huizi Mao, Yu~Wang, and Bill Dally.
\newblock Deep gradient compression: Reducing the communication bandwidth for
  distributed training.
\newblock In \emph{{ICLR}}, 2018.

\bibitem[Liu et~al.(2020)Liu, Cao, Yoshikawa, and Chen]{Liu_2020}
Ruixuan Liu, Yang Cao, Masatoshi Yoshikawa, and Hong Chen.
\newblock Fedsel: Federated sgd under local differential privacy with top-k
  dimension selection.
\newblock \emph{Lecture Notes in Computer Science}, 2020.

\bibitem[Makadia and Ryan(2014)]{Makadia}
Rupa Makadia and Patrick~B. Ryan.
\newblock Transforming the premier perspective® hospital database into the
  observational medical outcomes partnership (omop) common data model.
\newblock In \emph{EGEMS}, 2014.

\bibitem[Mcdonald et~al.(2012)Mcdonald, Peng, Sridharan, Foust, Kogan, Pezzin,
  and Feldman]{MRCI}
Margaret Mcdonald, Timothy Peng, Sridevi Sridharan, Janice Foust, Polina Kogan,
  Liliana Pezzin, and Penny Feldman.
\newblock Automating the medication regimen complexity index.
\newblock \emph{Journal of the American Medical Informatics Association :
  JAMIA}, 2012.

\bibitem[McMahan et~al.(2016)McMahan, Moore, Ramage, Hampson, and
  y~Arcas]{FedAVG}
H.~Brendan McMahan, Eider Moore, Daniel Ramage, Seth Hampson, and
  Blaise~Ag{\"u}era y~Arcas.
\newblock Communication-efficient learning of deep networks from decentralized
  data.
\newblock In \emph{AISTATS}, 2016.

\bibitem[McMahan et~al.(2018)McMahan, Ramage, Talwar, and
  Zhang]{Client-DP-McMahan}
H.~Brendan McMahan, Daniel Ramage, Kunal Talwar, and Li~Zhang.
\newblock Learning differentially private recurrent language models.
\newblock In \emph{International Conference on Learning Representations}, 2018.

\bibitem[Melis et~al.(2018)Melis, Song, Cristofaro, and
  Shmatikov]{Property_inference}
Luca Melis, Congzheng Song, Emiliano~De Cristofaro, and Vitaly Shmatikov.
\newblock Inference attacks against collaborative learning.
\newblock 2018.

\bibitem[Mironov et~al.(2019)Mironov, Talwar, and Zhang]{MironovTZ19}
Ilya Mironov, Kunal Talwar, and Li~Zhang.
\newblock R{\'{e}}nyi differential privacy of the sampled gaussian mechanism.
\newblock 2019.

\bibitem[More(2016)]{more2016survey}
Ajinkya More.
\newblock Survey of resampling techniques for improving classification
  performance in unbalanced datasets.
\newblock 2016.

\bibitem[Nasr et~al.(2019)Nasr, Shokri, and Houmansadr]{NasrSH19}
Milad Nasr, Reza Shokri, and Amir Houmansadr.
\newblock Comprehensive privacy analysis of deep learning: Passive and active
  white-box inference attacks against centralized and federated learning.
\newblock In \emph{{IEEE} Symposium on Security and Privacy}, 2019.

\bibitem[Oliphant(2006)]{Numpy}
Travis~E Oliphant.
\newblock \emph{A guide to NumPy}, volume~1.
\newblock Trelgol Publishing USA, 2006.

\bibitem[Rajkomar and al.(2018)]{rajkomar-npj18}
Alvin Rajkomar and al.
\newblock Scalable and accurate deep learning with electronic health records.
\newblock \emph{npj Digital Medicine}, 2018.

\bibitem[Rothchild et~al.(2020)Rothchild, Panda, Ullah, Ivkin, Stoica,
  Braverman, Gonzalez, and Arora]{rothchild2020fetchsgd}
Daniel Rothchild, Ashwinee Panda, Enayat Ullah, Nikita Ivkin, Ion Stoica,
  Vladimir Braverman, Joseph Gonzalez, and Raman Arora.
\newblock Fetchsgd: Communication-efficient federated learning with sketching,
  2020.

\bibitem[Seide et~al.(2014)Seide, Fu, Droppo, Li, and Yu]{SeideFDLY14}
Frank Seide, Hao Fu, Jasha Droppo, Gang Li, and Dong Yu.
\newblock 1-bit stochastic gradient descent and its application to
  data-parallel distributed training of speech dnns.
\newblock In \emph{{INTERSPEECH}}, 2014.

\bibitem[Shokri and Shmatikov(2015)]{ShokriS15}
Reza Shokri and Vitaly Shmatikov.
\newblock Privacy-preserving deep learning.
\newblock In \emph{{CCS}}, 2015.

\bibitem[Wang et~al.(2018)]{ATOMO}
Hongyi Wang et~al.
\newblock Atomo: Communication-efficient learning via atomic sparsification.
\newblock In \emph{NeurIPS}, 2018.

\bibitem[Wen and al.(2017)]{TERNGRAD}
Wei Wen and al.
\newblock Terngrad: Ternary gradients to reduce communication in distributed
  deep learning.
\newblock 2017.

\bibitem[Xiao et~al.(2017)Xiao, Rasul, and Vollgraf]{Fashion-MNIST}
Han Xiao, Kashif Rasul, and Roland Vollgraf.
\newblock Fashion-mnist: a novel image dataset for benchmarking machine
  learning algorithms.
\newblock 2017.

\bibitem[Zhao et~al.(2020)Zhao, Mopuri, and Bilen]{idlg}
Bo~Zhao, Konda~Reddy Mopuri, and Hakan Bilen.
\newblock idlg: Improved deep leakage from gradients.
\newblock 2020.

\bibitem[Zhu et~al.(2019)Zhu, Liu, and Han]{ZhuLH19}
Ligeng Zhu, Zhijian Liu, and Song Han.
\newblock Deep leakage from gradients.
\newblock In Hanna~M. Wallach, Hugo Larochelle, Alina Beygelzimer, Florence
  d'Alch{\'{e}}{-}Buc, Emily~B. Fox, and Roman Garnett, editors, \emph{NeurIPS
  2019}, 2019.

\bibitem[Zhu et~al.(2020)Zhu, Yu, Tsai, Pittaluga, Faraki, chandraker, and
  Wang]{zhu2020votingbased}
Yuqing Zhu, Xiang Yu, Yi-Hsuan Tsai, Francesco Pittaluga, Masoud Faraki,
  Manmohan chandraker, and Yu-Xiang Wang.
\newblock Voting-based approaches for differentially private federated
  learning, 2020.

\end{thebibliography}

\clearpage
\appendix

\section{Medical data: Data pre-processing \& experimental setup details}
\label{sec:app}

This section describes our medical dataset and the experimental setting which is used to evaluate the accuracy and the privacy of our proposals.

\subsection{Medical Dataset}

\label{sec:medical_dataset_desc}

\subsubsection{The In-hospital Mortality Prediction Scenario}

The ability to accurately predict the risks in the patient's perspectives of evolution is a crucial prerequisite in order to adapt the care that certain patients receive \citep{Papier_Amela}.

We consider the scenario where several hospitals are collaborating to train models for in-hospital mortality prediction using
our Federated Learning schemes. 
This well-studied real-world problem consists in trying to precisely identify the patients who are at risk of dying from complications during their hospital stay \citep{Avati2018,rajkomar-npj18,Papier_Amela}. As commonly found in the literature \citep{Papier_Amela}, for such predictions, we focus on hospital admissions of adults hospitalized for at least 3 days, excluding elective admissions.


\subsubsection{The Premier Healthcare Database}

We used EHR data from the Premier healthcare database\footnote{\href{https://www.premierinc.com/newsroom/education/premier-healthcare-database-whitepaper}{https://www.premierinc.com/newsroom/education/premier-healthcare-database-whitepaper}} which is one of the largest clinical databases in the United States, collecting information from millions of patients over a period of 12 months from 415 hospitals in the USA \citep{Papier_Amela}. These hospitals are supposedly representative of the United States hospital experience \citep{Papier_Amela}. Each hospital in the database provides discharge files that are dated records of all billable items (including therapeutic and diagnostic procedures, medication, and laboratory usage) which are all linked to a given patient's admission \citep{Papier_Amela,Makadia}. 

The initial snapshot of the database used in our work (before pre-processing step) comprises the EHR data of 1,271,733 hospital admissions.
Electronic Health Record (EHR) is a digital version of a patient’s paper chart readily available in hospitals. For developing supervised learning and specifically deep learning models, we focus on a specific set of features from EHR data. The features of interest that capture the patients information are summarized in Table~\ref{tab:data_description}. There is a total of 24,428 features per patient, mainly due to the variety of drugs possibly served. As in \cite{Avati2018}, we also removed all the features which appear on less than 100 patients' records, hence, the number of features was reduced to 7,280 features. 

The Medication regimen complexity index (MRCI) \citep{MRCI} is an aggregate score computed from a total of 65 items, whose purpose is to indicate the complexity of the patient's situation. The minimum MRCI score for a patient is 1.5, which represents a single tablet or capsule taken once a day as needed (single medication). However the maximum is not defined since the number of medications increases the score \citep{MRCI}. In our case, after statistical analysis of our dataset, we consider the MRCI score as ranging from 2 to 60.

Most real datasets like ours are generally imbalanced with a skewed distribution between the classes. In our case, the positive cases (patients who die during their hospital stay) represent only 3\% of all patients. Table \ref{tab:class_prop} gives more details about this distribution after the pre-processing step which is discussed in \ref{preprocessing}. To deal with this well-known problem, we have decided to use downsampling technique \citep{more2016survey,he2009learning}, a standard solution used for this purpose, as used in \cite{our_cs}.

\subsection{Preprocessing} 

\label{preprocessing}

\begin{enumerate}
    \item \textbf{Features normalization}: we extract from the dataset the values of each feature represented in Table \ref{tab:data_description}. For gender, we use one-hot encoding: Male, Female and Unknown. Similarly, for admission type we use 4 features: Emergency, Urgent, Trauma Center, and Unknown \footnote{\url{https://www.resdac.org/cms-data/variables/claim-inpatient-admission-type-code-ffs}}. For drugs, we extract 24,419 features which correspond to the different drugs (name and dosage). A given patient receives only a few of the possible drugs served, resulting in a very sparse patient's record. We use a MinMax normalization for age and MRCI in order to rescale the values of these features between 0 and 1 (using MinMaxScaler class of scikit-learn\footnote{\url{https://scikit-learn.org/stable/modules/generated/sklearn.preprocessing.MinMaxScaler.html}}). The labels that we consider are boolean: true means that the patient died during his hospital stay while false means she survived.

    \item \textbf{Patients filtering}: We consider patient and drug information of the first day at the hospital so that we can make predictions 24 hours after admission (as commonly found in the literature \citep{rajkomar-npj18,Papier_Amela}). We filter out the pregnant and new-born patients because the medication types and admission services are not the same for theses two categories of patients. Our model prediction is built without patients' historical medical data. This has the advantage to require minimum patient's information and to work for new patients.
    
    \item \textbf{Hospitals filtering}: The dataset contains 415 hospitals for a total size of 1,271,733 records. We split randomly the dataset into disjoint training and testing data (80\% and 20\% respectively).  The final dataset for testing contains 254,347 patients, with 7,882 deceased patients and 246,465 non-deceased patients (see Table~\ref{tab:class_prop}).
    
    Using Client-Level differential privacy requires to add more noise than Record-Level differential privacy, because the privacy purposes are not the same as detailled in Section~\ref{sec:backg}. To reduce the noise (when $\epsilon$ is fixed) and then improve the utility, we have to reduce the number of iterations or to reduce the sampling probability which are the parameters used to compute $\epsilon$. We therefore have two options to reduce the sampling probability:
    \begin{itemize}
        \item[-] Reducing the number of clients selected at each round $|\mathbb{K}|$. However this option also decreases the amount of data, and hence have a negative impact on the utility. We therefore preferred to use the next option.
        \item[-] Increasing the total number of clients $N$: we created more hospitals by splitting randomly the training data over 5010 "virtual" hospitals. We also, took care to have at least one in-hospital dead patient per hospital. Each hospital contains 203 patients. 356 patients are used as public dataset to define the \TOPK updated weights. We created 5010 hospitals in order to have approximately the same number of patients per hospital, each of them with some in-hospital dead patients. 
        
        In practise, Client-Level differential privacy is more adapted to an environment with a large set of clients as explained in \cite{Client-DP-McMahan,Client-DP-ETH-Zurich}.
    \end{itemize}

\end{enumerate}

\subsection{Imbalanced data}

The dataset of each hospital is imbalanced because the proportion of patients that leave the hospital alive is, fortunately, much larger than in-hospital dead patients. To deal with this well-known problem, we have decided to use downsampling technique \citep{more2016survey,he2009learning}, a standard solution used for this purpose. \footnote{We have also tested weighted loss function and oversampling techniques. But, we noticed experimentally that downsampling technique outperforms the other techniques for all the schemes.}



\subsection{Performance Metrics}

We use the following metrics:
\begin{itemize}
    \label{sec:metrics}
    \item \emph{Balanced accuracy}  \citep{Balanced_acc_1,Balanced_acc_2} is computed as  $1/2 \cdot (\frac{\TP}{\Pos}+\frac{\TN}{\Neg}) = \frac{\TPR+\TNR}{2}$ and is mainly used with imbalanced data. \emph{True Positive Rate} ($\TPR$) and \emph{True Negative Rate} ($\TNR$):
    $\TPR=\frac{\TP}{\Pos}$ and $\TNR=\frac{\TN}{\Neg}$, where $\Pos$ and $\Neg$ are the number of positive and negative instances, respectively, and $\TP$ and $\TN$ are the number of true positive and true negative instances.
    We note that traditional (``non-balanced'') accuracy metrics such as $\frac{\TP+\TN}{\Pos+\Neg}$ can be misleading for very imbalanced data \cite{akosa2017predictive}: in our dataset, the minority class has only 3\% of all the training samples (see Table~\ref{tab:class_prop}), which means that a biased (and totally useless) model always predicting the majority class would have a (non-balanced) accuracy of  $97\%$. 
    \item The \emph{area under the ROC curve} (\AUC) is also a frequently used accuracy metric. The ROC curve is calculated by varying the prediction threshold from 1 to 0, when \TPR and \FPR are calculated at each threshold. The area under this curve is then used to measure the quality of the predictions. A random guess has an $\AUC$ value of 0.5, whereas a perfect prediction has the largest $\AUC$ value of 1.  
\end{itemize}

\subsection{Evaluation Method.}
First, we split randomly the dataset of each hospital into disjoint training and testing data (80\% and 20\% respectively).
An entire federated run is executed with this split, and all the metrics are evaluated in every round on the union of all clients' testing data. 
All metric values of the round with the best balanced metric are recorded.

\subsubsection{Model architecture}

As in \cite{Avati2018,our_cs}, we use a fully connected neural network model with the following architecture: two hidden layers of 200 units, which use a Relu activation function followed by an output layer of 1 unit with sigmoid activation function and a binary cross entropy loss function. This results in 1,496,601 parameters in total. We tune $\eta$ from 0.01 to 0.5 with an increment value of 0.005. As in \cite{our_cs}, we fix the momentum parameter $\rho$ to 0.9 and the global learning rate $\eta_{G}$ to 1.0. The number of chunks is set to $P =100$ (refers to \cite{our_cs} for details). The hyperparameters used by each of the considered schemes are summarized in Table~\ref{tab:hyperparameters}.



\section{Fashion-mnist data: Data pre-processing \& experimental setup details}
\subsection{Data Description}
\label{sec:fashion_mnist_desc}
Fashion-MNIST database of fashion articles consists of 60,000 28x28 grayscale images of 10 fashion categories, along with a test set of 10,000 images \cite{Fashion-MNIST} \cite{KERAS_datasets}.

\subsection{Public data description}
\label{sec:mnist_desc}
The MNIST database of handwritten digits. It consists of 28 x 28 grayscale images of digit items and has 10 output classes. The training set contains 60,000 data samples while the test/validation set has 10,000 samples \cite{MNIST} \cite{KERAS_datasets}.


\subsection{Preprocessing}
The pixel of each image is an unsigned integer in the range between 0 and 255. We rescale them to the range [0,1] instead. \medskip

\subsection{Model architecture}
For Fashion-MNIST, we use a model \cite{FedAVG,our_cs} with the following architecture: a convolutional neural network (CNN) with two 5x5 convolution layers (the first with 32 filters, the second with 64, each followed with 2x2 max pooling), a fully connected layer with 512 units and ReLu activation, and a final softmax output layer. This results in 1,663,370 parameters in total. We tune $\eta$ from 0.01 to 0.5 with an increment value of 0.005. As in \cite{our_cs}, we fix the momentum parameter $\rho$ to 0.9 and the global learning rate $\eta_{G}$ to 0.35. Same for the number of chunks used $P =200$ (refers to \cite{our_cs} for more details).  The hyperparameters used by each of the considered schemes are summarized in Table~\ref{tab:hyperparameters}.


\section{Computational environment}

Our experiments were performed on a server running Ubuntu 18.04 LTS equipped with a Intel(R) Xeon(R) Silver 4114 CPU @ 2.20GHz, 192GB RAM, and two NVIDIA Quadro P5000 GPU card of 16 Go each. We use Keras 2.2.0 \cite{KERAS} with a TensorFlow  backend 1.12.0 \cite{TensorFlow} and Numpy 1.14.3 \cite{Numpy} to implement our models and experiments. We use  Python 3.6.5 and our code runs on a Docker container to simplify reproducibility. 

\section{Further experiments}
\label{sec:furthmore_experiments}
The goal of this section is to compare the performance of our proposed schemes FL-TOP and FL-TOP-DP with several baselines  according to different compression ratios. 
More specifically, we consider the following additional baselines:
\begin{itemize}
    \item FL-BAS-2: As in FL-BASIC, only a randomly selected set of parameters are selected and sent to the server at each round. Importantly, none of the parameters are reinitialized during training.
    \item FL-BAS-3: This baseline is the same as FL-BASIC, except that the set of random parameters is fixed over all the rounds.
    \item FL-BAS-4: Same as FL-BAS-2, except that the set of random parameters is the same over all the rounds.
    \item FL-TOP-BIS: Similarly to FL-TOP, it uses the same \TOPK parameters over the whole training. The only difference is that the $n-K$ non-\TOPK parameters are not re-initialized after each SGD iteration. As in FL-TOP, after $\Tgd$ SGD iterations, clients send the update of the \TOPK parameters to the server.
\end{itemize}

Note that all compression operators in the new baselines are still linear (just like FL-TOP-DP), and hence they can also be used with secure aggregation. Their private extensions (i.e., FL-BAS-2-DP, FL-BAS-3-DP, FL-BAS-4-DP and FL-TOP-BIS-DP) also clip and then noise the compressed updates as in FL-TOP-DP. The selection of sensitivity $S$ happens similarly to FL-TOP-DP and FL-BASIC-DP using the public data as described in Section \ref{sec:exp}.



\subsection{Results}

Table~\ref{tab:description_results_Fashion_MNIST} shows the best accuracy over 200 rounds for each scheme on the Fashion-MNIST dataset. $\mathit{Round}$ corresponds to the round when the best accuracy is achieved and $\mathit{Cost}$ is the average bandwidth consumption calculated as: $r \times n \times 32 \times \mathit{Round} \times C$, where 32 is the number of bits necessary to represent a float value, $n$ is the uncompressed model size, $r=\frac{|\mathbb{T}|}{n}$, $|\mathbb{T}|$ is the compressed model size, $C$ is the sampling probability of a client, and $\mathit{Round}$ is the round when we get the the best accuracy.

Table~\ref{tab:description_results_Medical_data_part_1} and Table~\ref{tab:description_results_Medical_data_part_2} display the best balanced accuracy over 100 rounds for each scheme on the Medical dataset. AUROC corresponds to the AUROC value when the best balanced accuracy is reached, $\mathit{Round}$ is the round when we get the best balanced accuracy, and finally, Cost is the average bandwidth consumption calculated as for the Fashion-MNIST dataset described above.

On the medical data (see Table~\ref{tab:description_results_Medical_data_part_1} and \ref{tab:description_results_Medical_data_part_2}), our schemes FL-TOP and FL-TOP-DP reach 0.64 of balanced accuracy and 0.70 of AUROC for $r=0.01\%$, while FL-TOP-Bis and FL-TOP-Bis-DP, which are the best baselines, have 8\% less of balanced accuracy and 10\% less of AUROC for identical compression ratios. Furthermore, for larger compression ratios, FL-TOP and FL-TOP-DP have similar results to that of FL-TOP-Bis and FL-TOP-Bis-DP. However, above $r=1\%$, FL-TOP outperforms FL-TOP-BIS. The same holds for FL-TOP-DP, which outperforms FL-TOP-Bis-DP when $r$ is more than $0.05\%$.  

On Fashion-MNIST, FL-TOP performs better than other schemes below $r=10\%$. For $r=10\%$,  FL-CS and FL-TOP have the same accuracy of 0.85. FL-TOP-DP is the best DP scheme independently of the compression ratio $r$. 

Notice the the larger the compression ratio $r$ is the smaller the performance gap between our schemes and the baselines FL-BAS-1, FL-BAS-3. The same holds for their DP counterparts. This is mainly due to the fact that the larger $r$ is the more likely that all schemes update the same \TOPK parameters. 

FL-CS and FL-CS-DP fail to improve their model accuracy when $r=0.01\%$ on the medical dataset. The same holds for FL-BAS-3-DP when $r=0.1\%$ on the Fashion-MNIST dataset.

On Fashion-MNIST, there is a decrease of accuracy for each of FL-TOP-DP, FL-TOP-BIS-DP and FL-CS-DP from $r=5\%$ to $r=10\%$. Indeed, as suggested in \cite{our_cs}, it may be due to the increase of sensitivity $S$ which will also increase the noise and therefore its negative impact on convergence.  

\begin{table*}[!h]
	\caption{Descriptions of features} 
	\label{tab:data_description}
	\scalebox{0.95}{
	\begin{tabular}{|r|l|}
		\hline
		     \emph{Features}    			            & \emph{Descriptions} 		                 \\
		\hline
		Age	                & 	 Value in the range of 15 and 89	  \\
		\hline
		Gender               &   Male, Female or Unknown  \\
		\hline
		Admission type     &   Emergency, Urgent, Trauma Center: visits to a trauma center/hospital or Unknown \\ 
		\hline
		MRCI               &   Medication regimen complexity index score (ranging from 2 to 60)   \\
		\hline
		\multirow{4}{*}{Drugs and ICD9 codes} &   Drugs given to the patient on the $1^{st}$ day of hospitalization. The ICD9 codes are composed \\ & of procedures  and diagnosis codes,  the first gives details about the medical procedures performed \\ & on the patient and the second about the doctor's  diagnosis of the patient.  There is a total of 24,419 \\ & possible drugs and ICD9 codes \citep{ICD9}.  \\
		\hline
	\end{tabular}}
\end{table*}

\begin{table}[h!]
	\caption{Number of instances for our case study. The Medical dataset contains in total 1,271,733 records.}
	\label{tab:class_prop}
	\centering
	\scalebox{0.8}{
	\begin{tabular}{ccccc}
		\hline
		Data & Positive cases         			                & Negative cases		                            & Ratio     & Total   \\
		\hline
		Train & 	      32,106  	                & 		     985,280                           &      3.16\%     &  1,017,386  \\
		\hline
		Test & 	        7,882	                & 		        246,465                        & 3.10\%          &  254,347 \\
		\hline
	\end{tabular}}
\end{table}

\begin{algorithm}[h]
\small
		\caption{FL-STD: Federated Learning \label{alg:fed_learn}}
	\DontPrintSemicolon
	{\bf Server:}\;
	\Indp Initialize common model $w_0$\;
	\For {$t=1$ \KwTo $\Tcl$}
	{
	    Select $\mathbb{K}$ clients uniformly at random \;
		\For {\textrm{each} client $k$ \textrm{in} $\mathbb{K}$}
		{	
			$\Delta \mbf{w}_t^k = \mathbf{Client}_k(\mbf{w}_{t-1})$\;
		}
		$\mbf{w}_{t} = \mbf{w}_{t-1} + \sum_{k} \frac{|D_k|}{\sum_j |D_j|} \Delta \mbf{w}_{t}^{k}$\;
	}
	\KwOut{Global model $\mbf{w}_t$}\;
	\Indm {\bf $\mathbf{Client}_{k}(\mbf{w}_{t-1}^k)$:}\;
	\Indp
	$\mbf{w}_{t}^k = \mathbf{SGD}(D_k, \mbf{w}_{t-1}^k, \Tgd)$\;
	\KwOut{Model update $(\mbf{w}_{t}^k- \mbf{w}_{t-1}^k)$} 
\end{algorithm}

\begin{algorithm}[h]
\small
	\caption{Stochastic Gradient Descent \label{alg:sgd}}
	\DontPrintSemicolon
	\KwIn{$D$ : training data, $\Tgd$ : local epochs, $\mathbf{w}$ : weights}  
    \For {$t=1$ \KwTo $\Tgd$}
	{
	    Select batch $\mathbb{B}$ from $D$ randomly\;
	    $\mbf{w} = \mbf{w} - \eta \nabla f(\mathbb{B}; \mbf{w})$\;
	}
    \KwOut{Model $\mbf{w}$} 
\end{algorithm}

\begin{algorithm}[h]
\small
		\caption{FL-STD-DP: Federated Learning with Client Privacy \label{alg:fl_std_dp}}
	\DontPrintSemicolon
	{\bf Server:}\;
	\Indp Initialize common model $w_0$\;
	\For {$t=1$ \KwTo $\Tcl$}
	{
	    Select $\mathbb{K}$ clients randomly \;
		\For {each client $k$ \textrm{in} $\mathbb{K}$}
		{	
			$\Delta \tilde{\mathbf{w}}_t^k = \mathbf{Client}_k(\mbf{w}_{t-1})$\;
		}
		$\mbf{w}_{t} = \mbf{w}_{t-1} + \frac{1}{|\mathbb{K}|} \sum_{k} \Delta \tilde{\mathbf{w}}_t^k $\;
	}
    \Indm {\bf $\mathbf{Client}_{k}(\mbf{w}_{t-1}^k)$:}\;
    \Indp
	$\Delta \mbf{w}_t^k = \mathbf{SGD}(D_k, \mbf{w}_{t-1}^{k}, \Tgd) - \mbf{w}_{t-1}^k$\;
	$\Delta \hat{\mbf{w}}_t^k = \Delta \mbf{w}_t^k / \max\left(1, \frac{||\Delta \mbf{w}_t^k||_2}{S}\right)$\;
    \KwOut{$\mathsf{Enc}_{K_k}(\mathcal{G}(\Delta \hat{\mbf{w}}_t^k, S \mathbf{I}\sigma /\sqrt{|K|}))$}
\end{algorithm}

\begin{table}[]
    \centering
    \scalebox{0.8}{
    \begin{tabular}{|c|c|}
        \hline
          \emph{Datasets}     &     \emph{Common Parameters}  \\
         \cline{1-2}
        \multirow{4}{*}{Fashion-MNIST dataset}  & $C=1/60$; $N=6000$; $\Tcl=200$; \\ & $\Tgd=5$;  $|\mathbb{B}|=10$;  $|D_k|=10$; $n=1,663,370$; \\ & $\delta=10^{-5}$;  $SGD(\eta=0.215)$;  $\eta_{G}=0.35$; \\ & $\rho=0.9$; $P=200$; $\sigma=1.54$; $T_{\mathsf{init}}=5$  \\
        \cline{1-2}
        \multirow{3}{*}{Medical dataset}  & $C=100/5010$; $N=5010$; $\Tcl=100$; $\Tgd=40$; \\ & $n=1,496,601$; $\delta=10^{-5}$; $SGD(\eta=0.1)$; $\eta_{G}=1.0$; \\ & $\rho=0.9$;  $P=100$;  $\sigma=1.49$; $T_{\mathsf{init}}=40$  \\
        \hline 
        
    \end{tabular}}
    \caption{Common environment between the schemes. $\rho$, $\eta_{G}$ and $P$ are only used with FL-CS and FL-CS-DP.}
    \label{tab:hyperparameters}
\vspace{-.3cm}
\end{table}


\begin{table}[!ht]
    \centering
    \scalebox{0.8}{
    \begin{tabular}{|c|c|c|c|c|c|}
        \hline
         \multirow{2}{*}{Algorithms} & \multicolumn{5}{c|}{\emph{Compression ratio ($r$)}} \\
         \cline{2-6}
         &   \emph{0.1\%} & \emph{0.5\%} & \emph{1\%} & \emph{5\%} & \emph{10\%} \\
         \hline
        FL-BASIC-DP    & 0.05 & 0.12 & 0.16 & 0.34 & 0.45 \\
        \cline{1-6}
        FL-BAS-2-DP    & 0.07 & 0.16 & 0.23 & 0.52 & 0.75 \\
        \cline{1-6}
        FL-BAS-3-DP    & 0.05  & 0.11 & 0.16 & 0.33 & 0.44 \\
        \cline{1-6}
        FL-BAS-4-DP    & 0.06 & 0.15 & 0.21 & 0.51 & 0.74 \\
        \cline{1-6}
        FL-CS-DP    & 0.21 & 0.26 & 0.32 & 0.57 & 0.79 \\
        \cline{1-6}
        FL-TOP-BIS-DP  & 1.25 & 1.59 & 1.79 & 2.18 & 2.34 \\
        \cline{1-6}
        FL-TOP-DP    & 0.50 & 0.61 & 0.64 & 0.87 & 1.0 \\
        \hline 
        
    \end{tabular}}
    \caption{Sensitivity S used for each scheme and for different compression ratio r on Fashion-MNIST. For FL-STD-DP, S is set to 2.40.}
    \label{tab:Sensitivity_fashion_mnist}
\vspace{-.3cm}
\end{table}

\begin{table}[!ht]
    \centering
    \scalebox{0.8}{
    \begin{tabular}{|c|c|c|c|c|c|c|c|}
        \hline
         \multirow{2}{*}{Algorithms} & \multicolumn{7}{c|}{\emph{Compression ratio ($r$)}} \\
         \cline{2-8}
         &   \emph{0.01\%} & \emph{0.05\%}  & \emph{0.1\%} & \emph{0.5\%} & \emph{1\%} & \emph{5\%} & \emph{10\%} \\
         \hline
        FL-BASIC-DP  & 0.01 & 0.03 & 0.05 & 0.11 & 0.16 & 0.34 & 0.46 \\
        \cline{1-8}
        FL-BAS-2-DP  & 0.01 & 0.03 & 0.04 & 0.09 & 0.14 & 0.31 & 0.44 \\
        \cline{1-8}
        FL-BAS-3-DP  & 0.01 & 0.04 & 0.06 & 0.12 & 0.18 & 0.35 & 0.49 \\
        \cline{1-8}
        FL-BAS-4-DP  & 0.02 & 0.03 & 0.05 & 0.12 & 0.15 & 0.31 & 0.44 \\
        \cline{1-8}
        FL-CS-DP  & 0.002 & 0.005 & 0.006 & 0.01 & 0.02 & 0.04 & 0.06 \\
        \cline{1-8}
        FL-TOP-BIS-DP  &  0.60 & 0.73 & 0.81 & 1.03 & 1.13 & 1.31 & 1.32 \\
        \cline{1-8}
        FL-TOP-DP  & 0.23 & 0.46 & 0.59 & 1.03 & 1.18 & 1.31 & 1.32 \\
        \hline 
        
    \end{tabular}}
    \caption{Sensitivity S used for each scheme and for different compression ratio r on the medical dataset. For FL-STD-DP, S is set to 1.40.}
    \label{tab:Sensitivity_medical_data}
\vspace{-.3cm}
\end{table}

\begin{table*}[!ht]
    \centering
    \scalebox{0.7}{
    \begin{tabular}{|c|c|c|c|c|c|c|}
        \hline
         \multirow{2}{*}{\emph{Compression ratio ($r$)}} & \multirow{2}{*}{Algorithms} & \multicolumn{5}{c|}{\emph{Performance}} \\
         \cline{3-7}
         & &  \emph{Accuracy}  & \emph{Round} & \emph{Downstream Cost } (Kilobyte) & \emph{Upstream Cost} (Kilobyte) & \emph{$\epsilon$} \\
         \hline

        \multirow{14}{*}{$0.1\%$} &  FL-BASIC   &  0.14 & 111 & 12308.94 & 12.31 & N/A  \\
        \cline{2-7}
        &  FL-BAS-2  &  0.16  & 185 & 20514.9 & 20.51 & N/A  \\
        \cline{2-7}
        &  FL-BAS-3  &  0.27  & 200 & 22.17 & 22.17 & N/A  \\
        \cline{2-7}
        &  FL-BAS-4  &  0.17  & 200 & 22.17 & 22.17 & N/A  \\
        \cline{2-7}
        &  FL-CS  & 0.37   & 200 & 22178.27 & 22.17 & N/A  \\
        \cline{2-7}
        &  FL-TOPK-BIS  &  0.59  & 198 & 21.95 & 21.95 & N/A  \\
        \cline{2-7}
        &  FL-TOP   & \textbf{0.78}  & 199 & \textbf{22.06} & \textbf{22.06} &N/A  \\
        \cline{2-7}
        \cline{2-7}
        &  FL-BASIC-DP   &  0.14  & 167 & 18518.85 & 18.51 & 0.95 \\
        \cline{2-7}
        &  FL-BAS-2-DP  &  0.14  & 124 & 13750.53 & 13.75 & 0.88 \\
        \cline{2-7}
        &  FL-BAS-3-DP  &  - & - & - & - & -\\
        \cline{2-7}
        &  FL-BAS-4-DP  &  0.15  & 137 & 15.19 & 15.19 & 0.90 \\
        \cline{2-7}
        &  FL-CS-DP  &  0.36  & 197 & 21845.59 & 21.84 & 1\\
        \cline{2-7}
        &  FL-TOPK-BIS-DP  &  0.59  & 196 & 21.73 & 21.73 & 0.99 \\
        \cline{2-7}
        &  FL-TOP-DP   & \textbf{0.76}  & 199 &  \textbf{22.06} & \textbf{22.06} & \textbf{1} \\
        \hline 
        \hline
        \multirow{14}{*}{$0.5\%$} &  FL-BASIC   &  0.65  & 193 & 21402.03 & 107 & N/A  \\
        \cline{2-7}
        &  FL-BAS-2  &  0.46  & 196 & 21734.70 & 108.66 & N/A  \\
        \cline{2-7}
        &  FL-BAS-3  &  0.73  & 200 & 110.88 & 110.88 & N/A  \\
        \cline{2-7}
        &  FL-BAS-4  & 0.41  & 197 & 109.22 & 109.22 & N/A  \\
        \cline{2-7}
        &  FL-CS  &  0.57  & 185 & 20514.9 & 102.56 & N/A  \\
        \cline{2-7}
        &  FL-TOPK-BIS  &  0.76  & 200 & 110.88 & 110.88 & N/A  \\
        \cline{2-7}
        &  FL-TOP   & \textbf{0.82}  & 200 & \textbf{110.88} & \textbf{110.88} & N/A \\
        \cline{2-7}
        \cline{2-7}
        &  FL-BASIC-DP   &  0.59  & 200 & 22178.27 & 110.88 & 1\\
        \cline{2-7}
        &  FL-BAS-2-DP  &  0.38 & 200  & 22178.27 & 110.88 & 1\\
        \cline{2-7}
        &  FL-BAS-3-DP  & 0.56   & 200 & 110.88 & 110.88 & 1\\
        \cline{2-7}
        &  FL-BAS-4-DP  &  0.33  & 200 & 110.88 & 110.88 & 1\\
        \cline{2-7}
        &  FL-CS-DP  &  0.53  & 200 & 22178.27 & 110.88 & 1\\
        \cline{2-7}
        &  FL-TOPK-BIS-DP  &  0.68  & 184 & 102.01 & 102.01 & 0.97\\
        \cline{2-7}
        &  FL-TOP-DP   & \textbf{0.81} & 200 & \textbf{110.88}  & \textbf{110.88} & \textbf{1}\\
        \hline 
        \hline
        \multirow{14}{*}{$1\%$} &  FL-BASIC   &  0.71  & 194 & 21512.92 & 215.12 & N/A  \\
        \cline{2-7}
        &  FL-BAS-2  &  0.59  & 200 & 22178.27 & 221.77 & N/A  \\
        \cline{2-7}
        &  FL-BAS-3  &  0.76  & 200 & 221.77 & 221.77 & N/A  \\
        \cline{2-7}
        &  FL-BAS-4  & 0.56  & 195 & 216.23 & 216.23 & N/A  \\
        \cline{2-7}
        &  FL-CS  &  0.69  & 200 & 22178.27 & 221.77 &N/A  \\
        \cline{2-7}
        &  FL-TOPK-BIS  &  0.79  & 197 & 218.45 & 218.45 & N/A  \\
        \cline{2-7}
        &  FL-TOP   & \textbf{0.83}  & 200 & \textbf{221.77} & \textbf{221.77} & N/A \\
        \cline{2-7}
        \cline{2-7}
        &  FL-BASIC-DP   &  0.65  & 197 & 21845.59 & 218.45 & 1\\
        \cline{2-7}
        &  FL-BAS-2-DP  & 0.62  & 198  & 21956.48 & 219.56 & 1\\
        \cline{2-7}
        &  FL-BAS-3-DP  &  0.66  & 198 & 219.56 & 219.56 & 1\\
        \cline{2-7}
        &  FL-BAS-4-DP  &  0.52  & 198 & 219.56 & 219.56 & 1\\
        \cline{2-7}
        &  FL-CS-DP  &  0.66  & 189 & 20958.46 & 209.58 & 0.98 \\
        \cline{2-7}
        &  FL-TOPK-BIS-DP  &  0.70  & 174 & 192.94 & 192.94 & 0.96\\
        \cline{2-7}
        &  FL-TOP-DP   & \textbf{0.81} & 183 & \textbf{202.92}  & \textbf{202.92} & \textbf{0.97}\\
        \hline
        \hline
        \multirow{14}{*}{$5\%$} &  FL-BASIC   &  0.78 & 196 & 21734.70 & 1086.73 & N/A  \\
        \cline{2-7}
        &  FL-BAS-2  &  0.72  & 199 & 22067.38 & 1103.36 & N/A  \\
        \cline{2-7}
        &  FL-BAS-3  &  0.81  & 199 & 1103.36 & 1103.36 &N/A  \\
        \cline{2-7}
        &  FL-BAS-4  &  0.76  & 196 & 1086.73 & 1086.73 & N/A  \\
        \cline{2-7}
        &  FL-CS  &  0.82  & 200 & 22178.27 & 1108.91 & N/A  \\
        \cline{2-7}
        &  FL-TOPK-BIS  &  0.83  & 196 & 1086.73 & 1086.73 & N/A  \\
        \cline{2-7}
        &  FL-TOP   & \textbf{0.84}  & 200 & \textbf{1108.91} & \textbf{1108.91} & N/A \\
        \cline{2-7}
        \cline{2-7}
        &  FL-BASIC-DP   &  0.76  & 195 & 21623.81 & 1081.18 & 0.99 \\
        \cline{2-7}
        &  FL-BAS-2-DP  &  0.72  & 195 & 21623.81 & 1081.18 & 0.99 \\
        \cline{2-7}
        &  FL-BAS-3-DP  &  0.76  & 199 & 1103.36 & 1103.36 & 1 \\
        \cline{2-7}
        &  FL-BAS-4-DP  &  0.75 & 191 & 1059.01 & 1059.01 & 0.99\\
        \cline{2-7}
        &  FL-CS-DP  &  0.78  & 160 & 17742.61 & 887.13 & 0.94 \\
        \cline{2-7}
        &  FL-TOPK-BIS-DP  &  0.71  & 152 & 842.77 & 842.77 & 0.92\\
        \cline{2-7}
        &  FL-TOP-DP   &  \textbf{0.81} & 152 & \textbf{842.77}  & \textbf{842.77} & \textbf{0.92}\\
        \hline 
        \hline
        \multirow{14}{*}{$10\%$} &  FL-BASIC   &  0.81  & 196 & 21734.70 & 2173.47 &N/A  \\
        \cline{2-7}
        &  FL-BAS-2  &    0.78  & 199 & 22067.38 & 2206.74 &N/A  \\
        \cline{2-7}
        &  FL-BAS-3  &  0.82  & 195 & 2162.38 & 2162.38 & N/A  \\
        \cline{2-7}
        &  FL-BAS-4  & 0.79   & 200 & 2217.83 & 2217.83 & N/A  \\
        \cline{2-7}
        &  FL-CS  &  0.85  & 182 & 20182.22 & 2018.22 & N/A  \\
        \cline{2-7}
        &  FL-TOPK-BIS  &   0.84 & 196 & 2173.47 & 2173.47 & N/A  \\
        \cline{2-7}
        &  FL-TOP   & \textbf{0.85}  & 199 & \textbf{2206.74} & \textbf{2206.74} &N/A \\
        \cline{2-7}
        \cline{2-7}
        &  FL-BASIC-DP   &  0.79  & 189 & 20958.46 & 2095.85 & 0.98\\
        \cline{2-7}
        &  FL-BAS-2-DP  &  0.77  & 189 & 20958.46 & 2095.85 & 0.98\\
        \cline{2-7}
        &  FL-BAS-3-DP  &  0.79  & 183 & 2029.31 & 2029.31 & 0.97 \\
        \cline{2-7}
        &  FL-BAS-4-DP  & 0.78  & 195 & 2162.38 & 2162.38 & 0.99\\
        \cline{2-7}
        &  FL-CS-DP  &  0.72  & 167 & 18518.85 & 1851.89 & 0.95\\
        \cline{2-7}
        &  FL-TOPK-BIS-DP  &  0.69  & 138 & 1530.30 & 1530.30 & 0.90\\
        \cline{2-7}
        &  FL-TOP-DP   &  \textbf{0.80} & 157 & \textbf{1740.99} & \textbf{1740.99} & \textbf{0.93}\\
        \hline 
        \hline

        \multirow{2}{*}{$100\%$} &  FL-STD  &  0.86  & 200 & 22178.27 & 22178.27 & N/A  \\
        \cline{2-7}
        &  FL-STD-DP  &  0.56 & 60 & 6653.48  & 6653.48 & 0.76 \\
        \hline
        
    \end{tabular}}
    \caption{Summary of results on Fashion-MNIST dataset.}
    \label{tab:description_results_Fashion_MNIST}
\vspace{-.3cm}
\end{table*}


\begin{table*}[!ht]
    \centering
    \scalebox{0.7}{
    \begin{tabular}{|c|c|c|c|c|c|c|c|}
        \hline
         \multirow{2}{*}{\emph{Compression ratio ($r$)}} & \multirow{2}{*}{Algorithms} & \multicolumn{6}{c|}{\emph{Performance}} \\
         \cline{3-8}
         & &   \emph{Bal\_Acc} & \emph{AUROC}  & \emph{Round} & \emph{Downstream Cost} (Kilobyte) & \emph{Upstream Cost} (Kilobyte) & \emph{$\epsilon$} \\
         \hline

        \multirow{14}{*}{$0.01\%$} &  FL-BASIC  & 0.49 & 0.45 & 100 & 11948.91 & 1.19 & N/A \\
        \cline{2-8}
        &  FL-BAS-2  & 0.49 & 0.45 & 94 & 11231.98 & 1.12 & N/A  \\
        \cline{2-8}
        &  FL-BAS-3 &  0.49 & 0.45 & 81 & 0.96 & 0.96 &N/A \\
        \cline{2-8}
        &  FL-BAS-4 &  0.49  & 0.49 & 100 & 1.19 & 1.19 &N/A \\
        \cline{2-8}
        &  FL-CS &   - & - & - & - & - &N/A \\
        \cline{2-8}
        &  FL-TOP-Bis & 0.59  & 0.63 & 100 & 1.19 & 1.19 &N/A \\
        \cline{2-8}
        &  FL-TOP & \textbf{0.64}  & \textbf{0.70} & 60 & \textbf{0.71} & \textbf{0.71} &N/A \\
        \cline{2-8}
        &  FL-BASIC-DP  & 0.49 & 0.45 & 6 & 716.93 & 0.07 & 0.74 \\
        \cline{2-8}
        &  FL-BAS-2-DP  & 0.49 & 0.45 & 100 & 11948.91 & 1.19 & 1\\
        \cline{2-8}
        &  FL-BAS-3-DP &  0.49  & 0.45 & 95 & 1.13 & 1.13 & 0.99\\
        \cline{2-8}
        &  FL-BAS-4-DP &  0.49  & 0.47 & 96 & 1.14 & 1.14 & 0.99\\
        \cline{2-8}
        &  FL-CS-DP &  -  & - & - & - & - & - \\
        \cline{2-8}
        &  FL-TOP-Bis-DP &  0.59  & 0.63 & 94 & 1.12 & 1.12  & 0.99\\
        \cline{2-8}
        &  FL-TOP-DP &  \textbf{0.64}  & \textbf{0.70} & 100 & \textbf{1.19} & \textbf{1.19} & \textbf{1}\\
        \hline 
        \hline
        \multirow{14}{*}{$0.05\%$} &  FL-BASIC  & 0.50 & 0.48 & 100 & 11948.91 & 5.97 & N/A \\
        \cline{2-8}
        &  FL-BAS-2  & 0.49 & 0.46 & 100 & 11948.91 & 5.97 & N/A  \\
        \cline{2-8}
        &  FL-BAS-3 &  0.51 & 0.49 & 100 & 5.97 & 5.97 & N/A \\
        \cline{2-8}
        &  FL-BAS-4 &  0.51  & 0.52 & 57 & 3.40 & 3.40 &N/A \\
        \cline{2-8}
        &  FL-CS &  0.51  & 0.50 & 100 & 11948.91 & 5.97 & N/A \\
        \cline{2-8}
        &  FL-TOP-Bis &  0.68  & 0.75 & 92 & 5.49 & 5.49 & N/A \\
        \cline{2-8}
        &  FL-TOP & \textbf{0.68}  & \textbf{0.75} & 54 & \textbf{3.22} & \textbf{3.22} &N/A \\
        \cline{2-8}
        &  FL-BASIC-DP  & 0.49 & 0.46 & 84 & 10037.08 & 5.02 & 0.96  \\
        \cline{2-8}
        &  FL-BAS-2-DP  & 0.49 & 0.46 & 100 & 11948.91 & 5.97 & 1 \\
        \cline{2-8}
        &  FL-BAS-3-DP &  0.50  & 0.48 & 99 & 5.91 & 5.91 & 1\\
        \cline{2-8}
        &  FL-BAS-4-DP &  0.52  & 0.51 & 100 & 5.97 & 5.97 & 1\\
        \cline{2-8}
        &  FL-CS-DP &  0.49  & 0.48 & 100 & 11948.91 & 5.97 & 1\\
        \cline{2-8}
        &  FL-TOP-Bis-DP &  0.68  & 0.75 & 92 & 5.49 & 5.49  & 0.98\\
        \cline{2-8}
        &  FL-TOP-DP &  \textbf{0.68}  & \textbf{0.75} & 99 & \textbf{5.91} & \textbf{5.91} & \textbf{1}\\
        \hline 
        \hline
        \multirow{14}{*}{$0.1\%$} &  FL-BASIC  & 0.51 & 0.51 & 99 & 11829.42 & 11.82 & N/A \\
        \cline{2-8}
        &  FL-BAS-2  & 0.50 & 0.47 & 100 & 11948.91 & 11.94 &N/A  \\
        \cline{2-8}
        &  FL-BAS-3 &  0.53  & 0.53 & 100 & 11.94 & 11.94 &N/A \\
        \cline{2-8}
        &  FL-BAS-4 &  0.50  & 0.53 & 94 & 11.23 & 11.23 &N/A \\
        \cline{2-8}
        &  FL-CS &  0.53  & 0.55 & 100 & 11948.91 & 11.94 &N/A \\
        \cline{2-8}
        &  FL-TOP-Bis &  0.69  & 0.76 & 100 & 11.94 & 11.94 &N/A \\
        \cline{2-8}
        &  FL-TOP & \textbf{0.69}  & \textbf{0.76} & 68 & \textbf{8.12} & \textbf{8.12} &N/A \\
        \cline{2-8}
        &  FL-BASIC-DP  & 0.50 & 0.49 & 100 & 11948.91 & 11.94 & 1  \\
        \cline{2-8}
        &  FL-BAS-2-DP  & 0.50 & 0.47 & 100 & 11948.91 & 11.94 & 1\\
        \cline{2-8}
        &  FL-BAS-3-DP &  0.55  & 0.56 & 100 & 11.94 & 11.94 & 1\\
        \cline{2-8}
        &  FL-BAS-4-DP &  0.51  & 0.52 & 100 & 11.94 & 11.94 & 1\\
        \cline{2-8}
        &  FL-CS-DP &  0.51  & 0.51 & 99 & 11829.42 & 11.82 & 1\\
        \cline{2-8}
        &  FL-TOP-Bis-DP & 0.68 & 0.75 & 89 & 10.63 & 10.63 & 0.98\\
        \cline{2-8}
        &  FL-TOP-DP &  \textbf{0.69}  & \textbf{0.76} & 85 & \textbf{10.15} & \textbf{10.15} & \textbf{0.97} \\
        \hline 
        \hline
        \multirow{14}{*}{$0.5\%$} &  FL-BASIC  & 0.58 & 0.68 & 100 & 11948.91 & 59.74 & N/A \\
        \cline{2-8}
        &  FL-BAS-2  & 0.56 & 0.58 & 99 & 11829.42 & 59.15 & N/A  \\
        \cline{2-8}
        &  FL-BAS-3 &  0.61  & 0.68 & 100 & 59.74 & 59.74 & N/A \\
        \cline{2-8}
        &  FL-BAS-4 &  0.56  & 0.59 & 100 & 59.74 & 59.74 &N/A \\
        \cline{2-8}
        &  FL-CS &  0.66  & 0.71 & 100 & 11948.91 & 59.74 & N/A \\
        \cline{2-8}
        &  FL-TOP-Bis &  0.71  & 0.78 & 100 & 59.74 & 59.74 &N/A \\
        \cline{2-8}
        &  FL-TOP & \textbf{0.71}  & \textbf{0.79} & 95 & \textbf{56.76} & \textbf{56.76} & N/A \\
        \cline{2-8}
        &  FL-BASIC-DP  & 0.57 & 0.64 & 100 & 11948.91 & 59.74   & 1\\
        \cline{2-8}
        &  FL-BAS-2-DP  & 0.57 & 0.59 & 100 & 11948.91 & 59.74  & 1\\
        \cline{2-8}
        &  FL-BAS-3-DP &  0.58  & 0.67 & 100 & 59.74 & 59.74 & 1\\
        \cline{2-8}
        &  FL-BAS-4-DP &  0.54  & 0.57 & 34 & 20.31 & 20.31 & 0.83\\
        \cline{2-8}
        &  FL-CS-DP &  0.61  & 0.68 & 100 & 11948.91 & 59.74 & 1\\
        \cline{2-8}
        &  FL-TOP-Bis-DP &  0.68  & 0.75 & 55 & 32.86 & 32.86 & 0.89\\
        \cline{2-8}
        &  FL-TOP-DP &  \textbf{0.69}  & \textbf{0.76} & 24 & \textbf{14.34} & \textbf{14.34} & \textbf{0.80}\\
        \hline
        
    \end{tabular}}
    \caption{Summary of results on Medical dataset (Part 1).}
    \label{tab:description_results_Medical_data_part_1}
\vspace{-.3cm}
\end{table*}

\begin{table*}[!ht]
    \centering
    \scalebox{0.7}{
    \begin{tabular}{|c|c|c|c|c|c|c|c|}
        \hline
         \multirow{2}{*}{\emph{Compression ratio ($r$)}} & \multirow{2}{*}{Algorithms} & \multicolumn{6}{c|}{\emph{Performance}} \\
         \cline{3-8}
         & &   \emph{Bal\_Acc} & \emph{AUROC}  & \emph{Round} & \emph{Downstream Cost} (Kilobyte) & \emph{Upstream Cost} (Kilobyte) & \emph{$\epsilon$} \\
         \hline
                \multirow{14}{*}{$1\%$} &  FL-BASIC  & 0.64 & 0.72 & 100 & 11948.91 & 119.49 & N/A \\
        \cline{2-8}
        &  FL-BAS-2  & 0.62 & 0.66 & 100 & 11948.91 & 119.49 &N/A  \\
        \cline{2-8}
        &  FL-BAS-3 &  0.62  & 0.66 & 85 & 101.57 & 101.57 &N/A \\
        \cline{2-8}
        &  FL-BAS-4 & 0.56 & 0.59 & 100 &  119.49 & 119.49 &N/A \\
        \cline{2-8}
        &  FL-CS &   0.68 & 0.75 & 100 & 11948.91 & 119.49 & N/A \\
        \cline{2-8}
        &  FL-TOP-Bis &  0.72  & 0.79 & 100 & 119.49 & 119.49 &N/A \\
        \cline{2-8}
        &  FL-TOP & \textbf{0.72}  & \textbf{0.79} & 58 & \textbf{69.30} & \textbf{69.30} &N/A \\
        \cline{2-8}
        &  FL-BASIC-DP  & 0.64 & 0.70 & 100 & 11948.91 & 119.49   & 1\\
        \cline{2-8}
        &  FL-BAS-2-DP  & 0.62 & 0.67 & 100 & 11948.91 & 119.49  & 1\\
        \cline{2-8}
        &  FL-BAS-3-DP &  0.61  & 0.71 & 100 & 119.49 & 119.49 & 1\\
        \cline{2-8}
        &  FL-BAS-4-DP &  0.57  & 0.66 & 100 & 119.49 & 119.49 & 1\\
        \cline{2-8}
        &  FL-CS-DP &  0.66  & 0.72 & 100 & 11948.91 & 119.49 & 1\\
        \cline{2-8}
        &  FL-TOP-Bis-DP &  0.68  & 0.74 & 53 & 63.33 & 63.33 & 0.89\\
        \cline{2-8}
        &  FL-TOP-DP &  \textbf{0.69}  & \textbf{0.76} & 22 & \textbf{26.29} & \textbf{26.29} & \textbf{0.79}\\
        \hline 
        \hline
        \multirow{14}{*}{$5\%$} &  FL-BASIC  & 0.72 & 0.80 & 100 & 11948.91 & 597.45 &N/A\\
        \cline{2-8}
        &  FL-BAS-2  & 0.68 & 0.75 & 100 & 11948.91 & 597.45 &N/A  \\
        \cline{2-8}
        &  FL-BAS-3 & 0.69 & 0.76 & 98 & 585.5 & 585.5 &N/A \\
        \cline{2-8}
        &  FL-BAS-4 &  0.66  & 0.72 & 100 & 597.45 & 597.45 &N/A \\
        \cline{2-8}
        &  FL-CS &   0.73 & 0.81 & 98 & 11709.93 & 585.5 & N/A \\
        \cline{2-8}
        &  FL-TOP-Bis &  0.72 & 0.79 & 100 & 597.45 & 597.45 &N/A \\
        \cline{2-8}
        &  FL-TOP &  \textbf{0.72}  & \textbf{0.80} & 95 & \textbf{567.57} & \textbf{567.57} &N/A \\
        \cline{2-8}
        &  FL-BASIC-DP  & 0.69 & 0.76 & 100 & 11948.91 & 597.45 & 1 \\
        \cline{2-8}
        &  FL-BAS-2-DP  & 0.68 & 0.75 & 98 & 11709.93 & 585.5 & 1 \\
        \cline{2-8}
        &  FL-BAS-3-DP &  0.65  & 0.71 & 90 & 537.70 & 537.70 & 0.98\\
        \cline{2-8}
        &  FL-BAS-4-DP &  0.67  & 0.74 & 98 & 585.5 & 585.5 & 1\\
        \cline{2-8}
        &  FL-CS-DP &  0.69  & 0.76 & 100 & 11948.91 & 597.45 & 1\\
        \cline{2-8}
        &  FL-TOP-Bis-DP &  0.67  & 0.74 & 38 & 227.03 & 227.03 & 0.84 \\
        \cline{2-8}
        &  FL-TOP-DP & \textbf{0.68} & \textbf{0.75} & 23 & \textbf{137.41} & \textbf{137.41} & \textbf{0.79}\\
        \hline 
        \hline
        \multirow{14}{*}{$10\%$} &  FL-BASIC  & 0.74 & 0.81 & 100 & 11948.91 & 1194.89 &N/A \\
        \cline{2-8}
        &  FL-BAS-2  & 0.70 & 0.77 & 100 & 11948.91 & 1194.89 &N/A  \\
        \cline{2-8}
        &  FL-BAS-3 &  0.72  & 0.80 & 98 & 1170.99 & 1170.99 &N/A \\
        \cline{2-8}
        &  FL-BAS-4 &  0.70  & 0.77 & 99 & 1182.94 & 1182.94 & N/A \\
        \cline{2-8}
        &  FL-CS &  0.74  & 0.82 & 100 & 11948.91 & 1194.89 & N/A \\
        \cline{2-8}
        &  FL-TOP-Bis &  0.72  & 0.80 & 100 & 1194.89 & 1194.89 &N/A \\
        \cline{2-8}
        &  FL-TOP & \textbf{0.74}  & \textbf{0.82} & 90 & \textbf{1075.40} & \textbf{1075.40} &N/A \\
        \cline{2-8}
        &  FL-BASIC-DP  & 0.69 & 0.76 & 99 & 11829.42 & 1182.94 & 1 \\
        \cline{2-8}
        &  FL-BAS-2-DP  & 0.69 & 0.76 & 95 & 11351.46 & 1135.15 & 0.99\\
        \cline{2-8}
        &  FL-BAS-3-DP &  0.69  & 0.76 & 95 & 1135.15 & 1135.15 & 0.99\\
        \cline{2-8}
        &  FL-BAS-4-DP &  0.69  & 0.76 & 100 & 1194.89 & 1194.89 & 1\\
        \cline{2-8}
        &  FL-CS-DP &  0.69  & 0.76 & 96 & 11470.95 & 1147.09 & 0.99\\
        \cline{2-8}
        &  FL-TOP-Bis-DP &  0.67  & 0.73 & 37 & 442.11 & 442.11 & 0.84\\
        \cline{2-8}
        &  FL-TOP-DP &  \textbf{0.68}  & \textbf{0.74} & 23 & \textbf{274.82} & \textbf{274.82} & \textbf{0.79}\\
        \hline 
        \hline

        \multirow{2}{*}{$100\%$} &  FL-STD  & 0.74 & 0.82 & 99 & 11829.42 & 11829.42 &N/A \\
        \cline{2-8}
        &  FL-STD-DP   & 0.66 & 0.72 & 62 & 7408.32 & 7408.32 & 0.91 \\
        \hline
        
    \end{tabular}}
    \caption{Summary of results on Medical dataset (Part 2).}
    \label{tab:description_results_Medical_data_part_2}
\vspace{-.3cm}
\end{table*}

\end{document}